\documentclass[conference]{IEEEtran}
\IEEEoverridecommandlockouts

\usepackage{cite}
\usepackage{amsmath,amssymb,amsfonts}
\usepackage{algorithmic}
\usepackage{graphicx}
\usepackage{textcomp}
\usepackage[dvipsnames]{xcolor}

\usepackage[hyphens]{url}
\usepackage{fancyhdr}
\usepackage[bookmarks=true,breaklinks=true,colorlinks,citecolor=blue,linkcolor=blue,urlcolor=blue]{hyperref}

\usepackage[normalem]{ulem}
\usepackage[final]{microtype}

\usepackage{subcaption}
\usepackage{listings}
\usepackage{cleveref}
\usepackage{multirow}
\usepackage{enumitem}
\usepackage{tikz}
\usepackage{soul}

\usepackage{def}

\def\BibTeX{{\rm B\kern-.05em{\sc i\kern-.025em b}\kern-.08em
    T\kern-.1667em\lower.7ex\hbox{E}\kern-.125emX}}

\title{Hardware-Assisted Virtualization of Neural Processing Units for Cloud Platforms}

\begin{document}

\bstctlcite{IEEEexample:BSTcontrol}



\author{\IEEEauthorblockN{Yuqi Xue}
\IEEEauthorblockA{University of Illinois Urbana-Champaign \\
yuqixue2@illinois.edu}\\
\IEEEauthorblockN{Lifeng Nai}
\IEEEauthorblockA{Google \\
lnai@google.com}
\and
\IEEEauthorblockN{Yiqi Liu}
\IEEEauthorblockA{University of Illinois Urbana-Champaign \\
yiqiliu2@illinois.edu}\\
\IEEEauthorblockN{Jian Huang}
\IEEEauthorblockA{University of Illinois Urbana-Champaign \\
jianh@illinois.edu}
}

\maketitle

\begin{abstract}
Cloud platforms today have been deploying hardware accelerators like neural processing units (NPUs) for powering machine learning (ML) 
inference services. To maximize the resource utilization while ensuring reasonable quality of service, a natural approach is to virtualize 
NPUs for efficient resource sharing for multi-tenant ML services. However, virtualizing NPUs for modern cloud platforms is not easy. 
This is not only due to the lack of system abstraction support for NPU hardware, but also due to 
the lack of architectural and ISA support for enabling fine-grained dynamic operator scheduling for virtualized NPUs. 

We present \pname{}, a holistic NPU virtualization framework. We investigate virtualization 
techniques for NPUs across the entire software and hardware stack. \pname{} consists of (1) a flexible NPU abstraction called 
vNPU, which enables fine-grained virtualization of the heterogeneous compute units in a physical NPU (pNPU); (2) a vNPU resource 
allocator that enables pay-as-you-go computing model and flexible vNPU-to-pNPU mappings for improved resource utilization and 
cost-effectiveness; (3) an ISA extension of modern NPU architecture for facilitating fine-grained tensor operator scheduling 
for multiple vNPUs. We implement \pname{} based on a production-level NPU simulator.
Our experiments 
show that \pname{} improves the throughput of ML inference services by up to 1.4$\times$ and reduces  
the tail latency by up to 4.6$\times$, while improving the NPU utilization by 1.2$\times$ on average, 
compared to state-of-the-art NPU sharing approaches.



\end{abstract}


\section{Introduction}
\label{sec:intro}

Machine learning (ML) is becoming the backbone for many popular ML services, such as online recommendation and natural language processing~\cite{mlaas:intro,
mlaas:industry,mlaas:aws, tpucloud}. 
To accelerate these ML services, cloud platforms have employed hardware accelerators like 
neural processing units (NPUs) as the mainstream compute engine~\cite{cloudtpu:google,ipu:graphcore,aws_inferentia,diannao,brainwave,tenstorrent}.

NPUs are highly specialized to accelerate the common operations in deep neural networks (DNNs), 
such as matrix multiplication and convolution.
A typical NPU device is a peripheral board with multiple NPU chips, 
and each chip has multiple NPU cores.
Each NPU core has matrix engines (\saname{s}) that leverage systolic arrays to perform matrix multiplications
and vector engines (\vuname{s}) for generic vector operations. 
A well-known example is the Google Cloud TPU~\cite{cloudtpu:google}.


A common approach to using NPUs in cloud platforms is to assign an entire NPU chip to a single ML application instance in a 
virtual machine (VM) or container via PCIe pass-through~\cite{tpucloud}. 
However, this disables resource sharing and causes severe resource underutilization of NPUs.
For instance, prior studies~\cite{v10:isca23}
disclosed 
that a majority of the DNN inference workloads cannot fully utilize TPU cores, due to their imbalanced demands on \saname{s} and \vuname{s}. 
Many DNN workloads
have diverse demands on the number of \saname{s} and 
\vuname{s} (see $\S$\ref{sec:npu_util_study}).
As a result, the one-size-fits-all approach is much less attractive for cloud platforms.   


To address the utilization challenge and ease the resource management for cloud platforms to accommodate diverse workload demands, it is desirable to virtualize hardware devices and enable resource sharing among multiple tenants. 
Unfortunately, 
modern cloud platforms have very limited virtualization support for NPUs across the software and hardware stack.

\vspace{0.2em}
\noindent
\underline{Lack of system abstraction support for NPUs.}
Unlike the system virtualization of multi-core processors~\cite{xen:sosp2003, adams:asplos2006}, 
NPUs have unique heterogeneous compute resources (i.e., \saname{s} and \vuname{s}). 
To circumvent this complexity, cloud platforms today expose homogeneous NPU cores to the user VMs. However, the existing abstraction at the NPU core level is too coarse-grained, as user workloads may have diverse resource requirements.
We need \textit{a flexible system abstraction that allows 
users to specify the \saname{}/\vuname{} resources} following the pay-as-you-go model~\cite{pay-as-you-go}.
Such an abstraction will simplify the NPU management for cloud platforms, including NPU resource (de)allocation, 
resource mapping, and scheduling. Prior studies investigated the system virtualization for 
FPGAs~\cite{fpgacloud,vital:asplos20,synergy:asplos21,mlvital:asplos21,amorphos:osdi18} and GPUs~\cite{gpucloud,micropreempt:osdi2022}.
However, they cannot be directly applied to NPUs, as they target different architectures. 

\vspace{0.2em}
\noindent
\underline{Lack of architectural support for NPU virtualization.}
Prior studies enabled the time-sharing of an NPU device at the task level, and support the preemption for 
prioritized tasks~\cite{baymax:asplos2016, prophet:asplos2017}. However, the coarse-grained time-sharing 
on the shared NPU board still suffers from severe resource underutilization, 
due to the lack of support of concurrent execution of multi-tenant workloads. 
Existing NPU sharing approaches either sacrifice isolation or suffer from
high preemption overhead~\cite{prema:hpca20}. 
V10~\cite{v10:isca23} enabled NPU sharing between multiple DNN workloads. However, it is still based on the 
time-sharing mechanism and suffers from operator interference between multi-tenant ML instances, 
resulting in poor performance isolation. 
As we move towards fine-grained NPU virtualization, 
we need \textit{architectural support to achieve both improved performance isolation and NPU utilization}.

\vspace{0.2em}
\noindent
\underline{Lack of ISA support for virtualized NPUs.}
To simplify the hardware design, NPUs commonly employ VLIW-style ISAs, and the ML compiler explicitly exploits the parallelism of the compute units~\cite{tpu,aichips:industry,amd:aiengine}. However, this requires the number of 
compute units to be explicitly specified at the compilation stage, and the number cannot be changed at runtime. In this case, the VLIW ISAs unnecessarily couple control flows of the compute units (i.e., \saname{s}). 
Even though some compute units of a shared NPU become available, they cannot be utilized by the active workload (except recompiling the DNN program).
This is caused by the fundamental tussle between dynamic scheduling and VLIW ISAs.
As the collocated ML instances have various demands on compute units at runtime, this limitation inevitably causes either NPU underutilization or performance interference.
We need to \textit{rethink the NPU ISA design to facilitate dynamic resource scheduling for virtualized NPUs}.

\vspace{0.2em}
Ideally, we wish to virtualize NPUs to enable flexible and fine-grained resource sharing and scheduling for improved NPU utilization and performance isolation. 
We present \pname{}, a hardware-assisted system virtualization framework for NPUs.

\vspace{0.2em}
\noindent
\underline{Our contributions.}
We first develop a simple yet flexible \textit{vNPU} abstraction. 
We use vNPU to create a virtualized NPU device for each ML instance. 
For each vNPU, the user can specify the number of different types of compute units (\saname{s}/\vuname{s}) 
on-demand or follow the pay-as-you-go model in cloud computing. 
We propose a new resource allocation mechanism that can decide the optimized vNPU configuration for different ML workloads, 
based on the analysis using ML compilers. As different ML services have various 
\saname{}/\vuname{} demands (see $\S$\ref{sec:motivation}), 
such an abstraction enables fine-grained resource allocation, which benefits both end users and cloud platform 
operators\footnote{The fine-grained resource allocation allows end users to allocate the NPU resources on demand, and enables 
cloud platforms to implement the pay-as-you-go model at a fine granularity as they have done for multi-core processors.}. 

\pname{} can map vNPUs to physical compute units of NPU cores in different manners,   
based on the service level objectives (SLOs) of ML services. To maximize the NPU utilization while ensuring performance isolation,
\pname{} enables fine-grained spatial sharing with resource harvesting. 
It also enables the oversubscription of NPU cores by temporally sharing \saname{}s/\vuname{}s among multiple vNPUs. 
Therefore, the idle compute units can be opportunistically utilized by collocated workloads.

To facilitate the dynamic scheduling for collocated vNPUs, \pname{} extends the VLIW-style ISA  
by reorganizing VLIW instructions into independent micro-Tensor operators (\utop{}s in $\S$\ref{sec:design}). \pname{} introduces
necessary architectural logic for fine-grained dynamic 
scheduling of \utop{}s on the shared physical NPU cores.
It allows one vNPU to harvest available compute cycles of \saname{s}/\vuname{s} from collocated vNPUs, without causing 
much interference. This is impossible with conventional VLIW-style ISAs, as they strictly couple the control flows of the (statically) allocated compute units. Our new architectural support enables \pname{} to offer 
the flexibility of NPU resource allocation and scheduling across the software (i.e., vNPU abstraction) and 
hardware (i.e., fine-grained \utop{} scheduling) stack. 
\pname{} requires minimum modifications to NPU chips (0.04\% die area cost) as well as ML compilers.

\begin{figure}[t]
    \centering
    \includegraphics[width=0.9\linewidth]{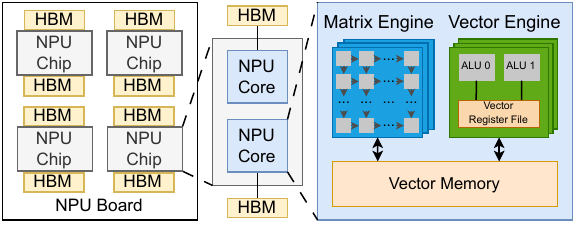}
    \caption{System architecture of a typical NPU board.}
    \label{fig:npu_arch}
\end{figure}

We implement \pname{} with a production-level NPU simulator following the typical TPU architecture.
We collect the traces of ML services as we run the MLPerf benchmarks~\cite{reddi2019mlperf} and the TPU reference models~\cite{tpu_supported_models} on the real 
Google TPUs. Our experiments with multi-tenant ML instances show that \pname{} can improve the throughput of ML inference services by up to 1.4$\times$ 
and reduce the tail latency by up to 4.6$\times$, while improving the NPU utilization by 1.2$\times$ on average, 
in comparison with state-of-the-art NPU sharing approaches. 
We summarize the contributions of \pname{} as follows:

\begin{itemize}[leftmargin=*]
\item We conduct a thorough study of DNN inference workloads on real NPU hardware, and investigate the NPU virtualization challenges within both 
	system and hardware stack ($\S$\ref{sec:motivation}).
\item We propose a new system abstraction named vNPU for enabling fine-grained virtualization of the heterogeneous compute units in NPU 
	cores ($\S$\ref{sec:vnpu_abstraction}). 
\item We present a new NPU resource allocation scheme and enable flexible vNPU-to-pNPU mappings  
	($\S$\ref{sec:vnpu_allocation} and $\S$\ref{sec:vnpu_mapping_hw}). 
\item We extend the VLIW-style ISAs and NPU architecture for enabling fine-grained dynamic scheduling of 
	virtualized NPUs for multi-tenant ML services ($\S$\ref{sec:isa_extension} and $\S$\ref{sec:arch_for_neuISA}).
\item We evaluate the efficiency and flexibility of our NPU virtualization framework with real-world DNN traces ($\S$\ref{sec:eval}). 
\end{itemize}

\begin{figure*}[t]
    \centering
    \includegraphics[width=0.9\linewidth]{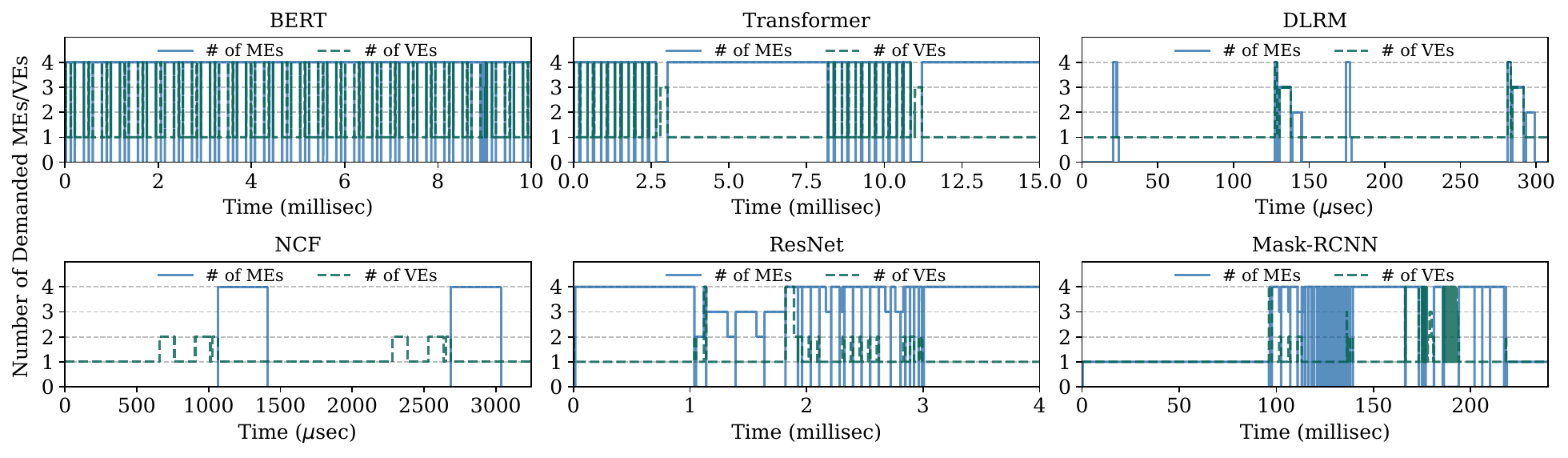}
    \caption{The number of \saname{}s and \vuname{}s demanded by DNN inference workloads over time (batch size = 8).}
    \label{fig:sa_vu_demand_ops}
\end{figure*}

\section{Background and Motivation}
\label{sec:motivation}

\subsection{NPU System Architecture}
\label{subsec:npuarch}

As shown in \Cref{fig:npu_arch}, an NPU board has multiple NPU chips, each chip has 
multiple NPU cores, each core is connected to an off-chip HBM. An NPU core has two types of compute units: 
matrix engines (\saname{}s) that perform matrix multiplications with systolic arrays; and vector engines (\vuname{}s) that perform generic vector operations. Each NPU core employs an on-chip SRAM to hide HBM access latency.
A typical example of NPU architecture in production is Google TPU~\cite{tpuarch:google:commACM20}. 

To run a DNN program on NPUs, ML compilers~\cite{tvm,PyTorch,tensorflow} generate a sequence of tensor operators, 
which are then translated into device-specific machine instructions.
An NPU core usually uses a VLIW-style ISA for simplifying the hardware. Each instruction contains multiple \saname{} slots, 
\vuname{} slots, load/store slots for accessing the SRAM, and other slots (e.g., for DMA operations with HBM).
The ML compilers can exploit the instruction-level parallelism with the knowledge of underlying compute resource,
such as the numbers of \saname{}s/\vuname{}s.

\begin{table}[t]
    \centering
    \caption{DNN models used as ML services in this paper. 
    }
    \begin{tabular}{|c|c|c|c|}
    \hline
        \multirow{2}{*}{\textbf{Category}} & \multirow{2}{*}{\textbf{Model Name}} & \multirow{2}{*}{\textbf{Abbrev.}} & \textbf{HBM Footprint} \\
        & & & {(batch size = 8)} \\\hline
	    Natural Language & BERT & BERT & 1.27GB \\
	    Processing & Transformer & TFMR & 1.54GB \\\hline
	    \multirow{2}{*}{Recommendation} & DLRM & DLRM & 22.38GB \\
	    & NCF & NCF & 11.10GB \\\hline
	    \multirow{3}{2cm}{Object Detection \& Segmentation} & Mask-RCNN & MRCNN & 3.21GB \\
	    & RetinaNet & RtNt & 860.51MB \\
	    & ShapeMask & SMask & 6.04GB \\\hline
	    \multirow{4}{*}{Image Classification} & MNIST & MNIST & 10.59MB \\
	    & ResNet & RsNt & 216.02MB \\
	    & ResNet-RS & RNRS & 458.17MB \\
	    & EfficientNet & ENet & 99.06MB \\\hline
    \end{tabular}
    \label{tab:workloads}
\end{table}

\subsection{Characterization of ML Inference Services}\label{sec:npu_util_study}

To motivate NPU virtualization, we conduct a study of resource demands of ML inference workloads and their impact on NPU utilization. 
We run various ML inference workloads from MLPerf benchmarks~\cite{reddi2019mlperf} and official TPU reference 
models~\cite{tpu_supported_models} (see \Cref{tab:workloads}), on a real Google TPUv4 board with 8 cores. 
Each core has four \saname{s} and two \vuname{s}. 
We profile the number of \saname{s}/\vuname{s} demanded by each workload with ML compiler techniques,  and the resource utilization with performance counters on the TPU core. 
We vary the batch size (8 by default). 
The HBM footprint of benchmarks ranges from 10.59MB to 22.38GB, which does not fully occupy the HBM on modern NPU chips (e.g., 32GB/96GB on TPUv4/TPUv5p~\cite{tpuv5p}).
We report the resource utilization on one TPU core,
as all cores perform identical computations with data parallelism. 

\begin{figure}[t]
    \centering
    \includegraphics[width=\linewidth]{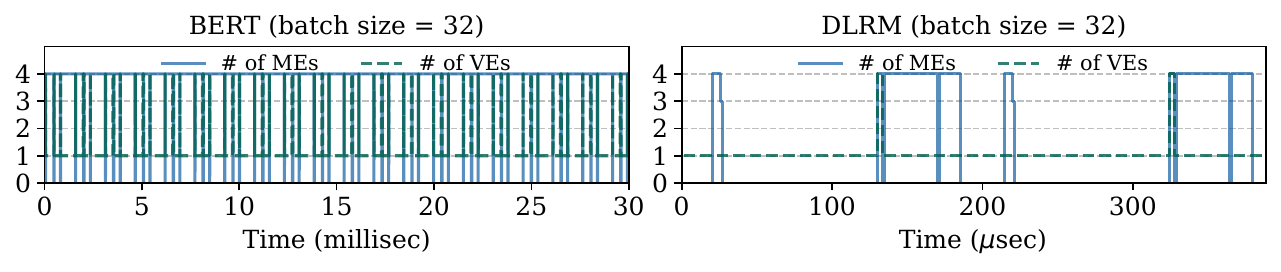}
    \caption{The number of \saname{}s and \vuname{}s demanded by DNN inference workloads with a larger batch size.}
    \label{fig:sa_vu_demand_batchsize}
\end{figure}

\begin{figure}[t]
    \centering
    \includegraphics[width=\linewidth]{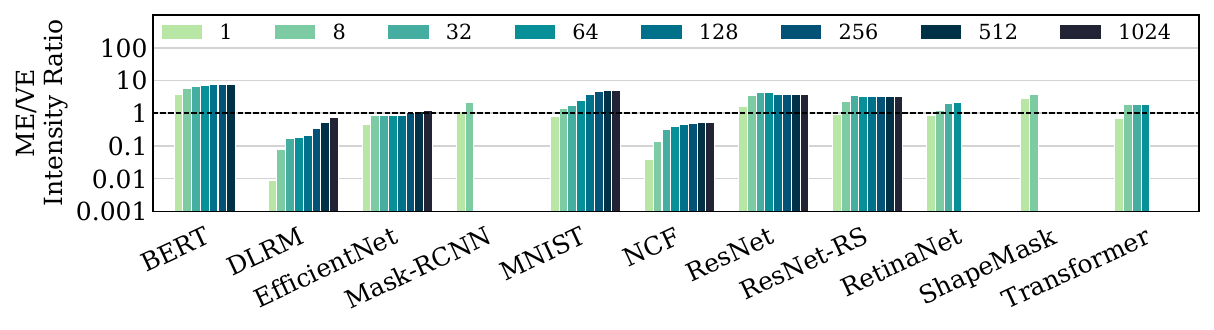}
	\caption{Intensity ratio of \saname{} vs. \vuname{} for different inference workloads (quantified by the 
	execution time of \saname{}/\vuname{}).}
    \label{fig:motiv_tpu_sa_vu_util}
\end{figure}


\begin{figure*}[t]
    \centering
    \includegraphics[width=0.9\linewidth]{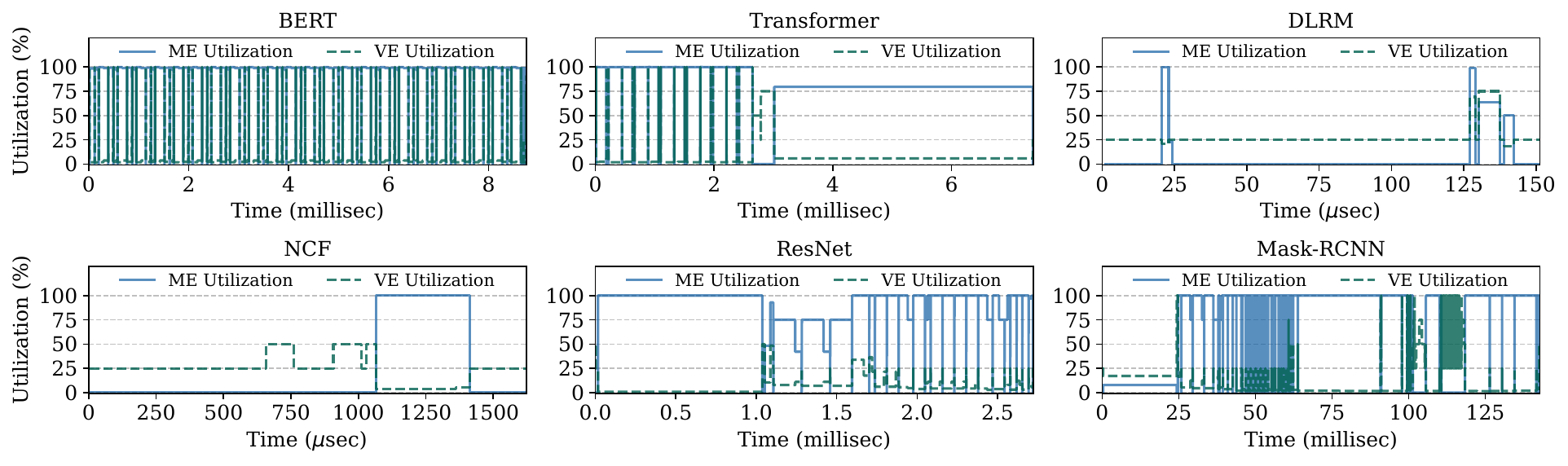}
    \caption{The utilization of \saname{} and \vuname{} of an inference request for representative DNN models.}
    \label{fig:sa_vu_util_ops}
\end{figure*}

\begin{figure}[t]
    \centering
    \includegraphics[width=\linewidth]{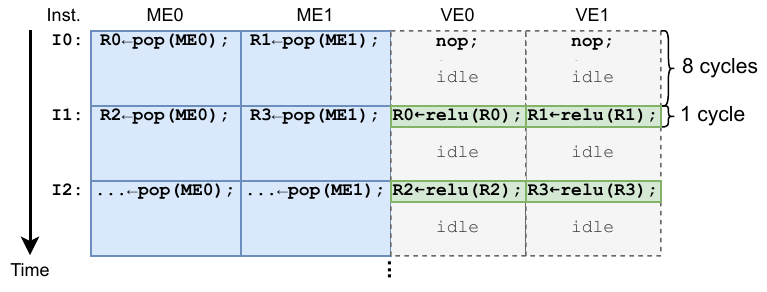}
    \caption{Example of \vuname{} underutilization in an \saname{}-intensive operator (fused matrix multiplication and ReLU activation).}
    \label{fig:sa_vu_underutil_op_example}
\end{figure}

\begin{figure}[t]
    \centering
    \includegraphics[width=\linewidth]{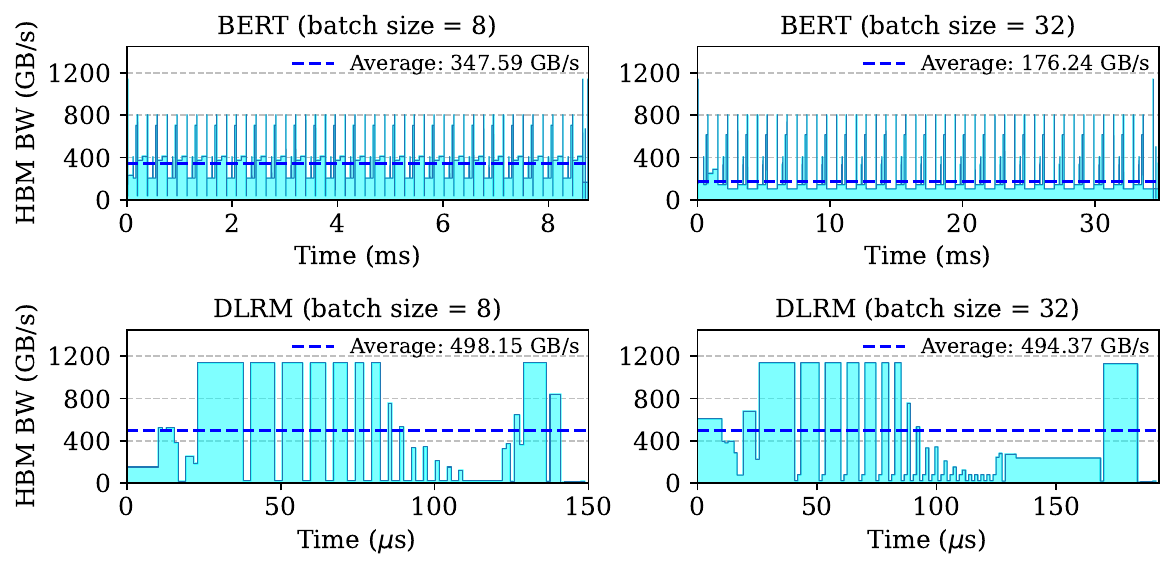}
    \caption{The HBM bandwidth utilization for representative DNN models with different batch sizes.}
    \label{fig:hbm_bw_util_ops}
\end{figure}




%

\vspace{0.2em}
\noindent
\textbf{Diverse demands on \saname{s}/\vuname{s}.}
An ML inference workload can have diverse resource demands over time, 
as different operators in a DNN model have vastly different demands on \saname{}s and \vuname{}s.
For each workload, we analyze the DNN execution graph generated by the ML compiler.
By default, the ML compiler picks the number of compute units for each operator to maximize the overall efficiency of the compute units based on the tensor shapes. We use this to quantify the \saname{}/\vuname{} demands.
Figure~\ref{fig:sa_vu_demand_ops} shows that DNN inference workloads have various \saname{}/\vuname{} demands over time. 
As we increase the batch size, 
we observe similar patterns (Figure~\ref{fig:sa_vu_demand_batchsize}).
Due to space limitations, we only show the results of BERT and DLRM. 
The imbalanced demands are determined by the ML model architecture.
For example, in~\Cref{fig:motiv_tpu_sa_vu_util}, 
ResNet is dominated by convolutions (\saname{}-intensive operators), while  
DLRM contains many vector operators, which do not utilize the \saname{} at all.
For workloads that cannot run with large batch sizes due to insufficient memory, we do not show 
them in \Cref{fig:motiv_tpu_sa_vu_util}.

\vspace{0.2em}
\noindent
\textbf{Low NPU resource utilization.}
The diverse demands on \saname{s}/\vuname{s} inevitably cause NPU underutilization. We quantify the 
percentage of idleness of the \saname{s}/\vuname{s} in \Cref{fig:sa_vu_util_ops}.
Although workloads like DLRM and NCF may appear to be \vuname{}-intensive, at least 20\% of their execution time 
still involves heavy \saname{} computation. For \saname{}-intensive models such as ResNet, many operators are also \vuname{}-intensive. To balance the demands on \saname{} and \vuname{}, the ML compiler can 
perform operator fusion to pipeline the execution of \saname{} and \vuname{}~\cite{xla, tvm, v10:isca23}. 
However, as such fusion opportunities are limited, most operators still have imbalanced \saname{}/\vuname{} 
demands after fusion.
\Cref{fig:sa_vu_underutil_op_example} shows an example of \vuname{} underutilization in an \saname{}-intensive operator.
Each \texttt{pop} operation takes 8 cycles to generate an $8\times 128$ output vector from the \saname{}, 
while each \vuname{} operation takes 1 cycle to post-process the output vector. As a result, the \vuname{} is idle for most of the time.

We also profile the HBM bandwidth utilization in Figure~\ref{fig:hbm_bw_util_ops}.
While the peak bandwidth almost reaches the hardware limit (1.2TB/s on a TPUv4 chip), the average bandwidth is as low as 176--498GB/s. This is because different operators in a DNN model have varying bandwidth demands. For example, in DLRM, the embedding lookup consumes high bandwidth, while the multi-layer perceptron (MLP) has low bandwidth requirements.
As we increase the batch sizes, the bandwidth consumption decreases for some workloads. For example, BERT 
is dominated by \saname{} operators, which become more compute-intensive with larger batch sizes; DLRM is \vuname{}-intensive, and 
\vuname{} operators have low compute intensity regardless of batch sizes. 
As some DNN operators underutilize the HBM bandwidth while other operators underutilize the compute resources, 
collocating DNN workloads on the same NPU core helps cloud platforms utilize both resources.

\subsection{NPU Virtualization: Challenges and Opportunities}
\label{subsec:challenges}
System virtualization offers the opportunity for supporting multi-tenancy and improving resource utilization. 
However, virtualizing NPUs suffers from unique challenges.

\vspace{0.2em}
\noindent
\textbf{New abstraction required for fine-grained virtualization.} 
As none of prior studies investigated NPU virtualization, it is unclear how the virtualized NPUs should be exposed to 
application instances. By virtualizing NPUs, we need to provide a simple yet effective abstraction, which can provide sufficient flexibility 
for users to specify the numbers of \saname{}s and \vuname{}s based on the workload demand and target SLOs (see $\S$\ref{sec:vnpu_allocation}). 
For instance, we should allocate more \saname{}s 
than \vuname{}s to an \saname{}-intensive workload, and vice versa.

However, even if we can allocate the most appropriate numbers of \saname{}s and \vuname{}s,  
the allocated resources still cannot be fully utilized, due to the diverse resource demands of different operators over time.
A static allocation of \saname{}s and \vuname{}s is insufficient. Instead, we need to enable dynamic resource scheduling.  
We should allow one workload 
to ``\textit{harvest}'' the underutilized compute units allocated to other workloads for improving the overall utilization of 
the NPU core and the Quality-of-Service (QoS) of collocated ML inference services. Unfortunately, current  
NPU architectures do not support such fine-grained resource scheduling and harvesting.  

\begin{figure}[t]
    \centering
    \includegraphics[width=\linewidth]{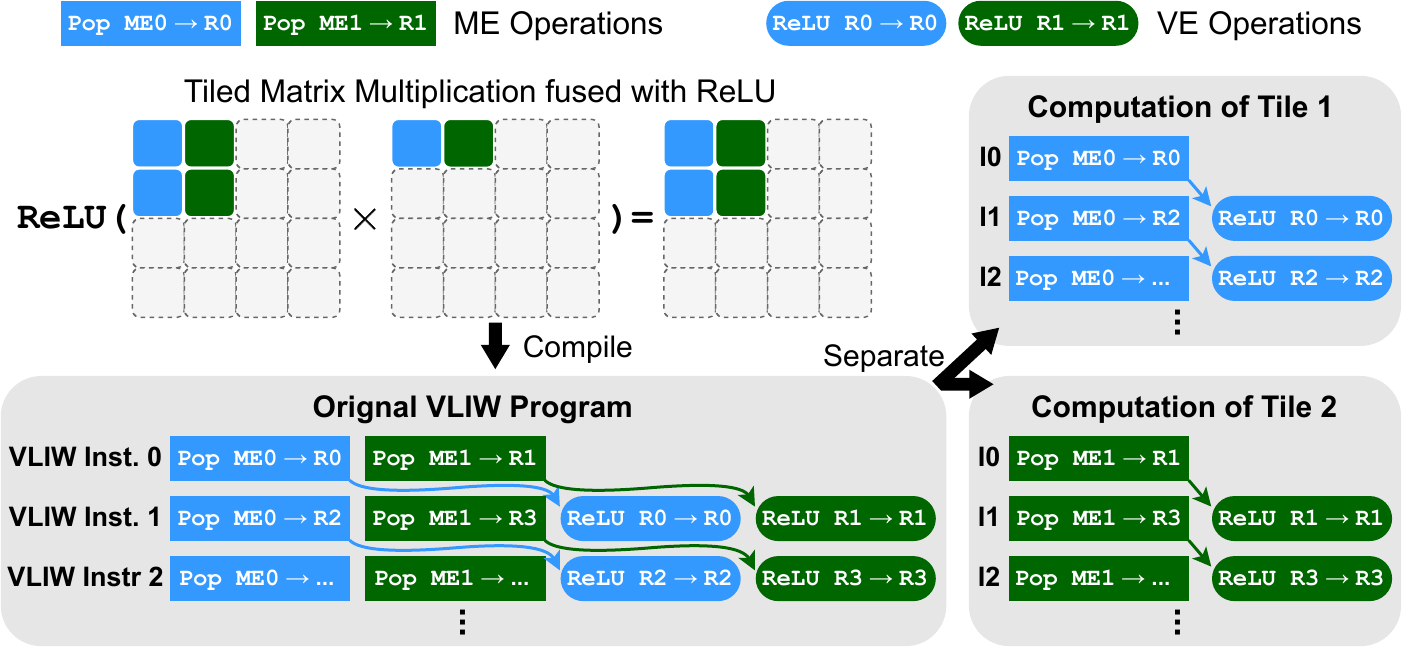}
    \caption{Execution of \saname{}s and \vuname{}s are separable. The arrows between instructions denote data dependencies.}
    \label{fig:control_flow_separation}
\end{figure}

\vspace{0.2em}
\noindent
\textbf{ISA limitations for enabling virtualized NPU scheduling.}
The fundamental limitations of modern NPU architectures prevent dynamic resource scheduling.  
To simplify the hardware design of NPUs, developers usually employ VLIW-style ISAs, and 
utilize ML compilers to exploit the instruction-level parallelism. However, the statically scheduled ISAs cannot 
fully exploit the hardware resources at runtime. 
They unnecessarily couple the control flows of all \saname{}s in a tensor operator, even though different \saname{}s can execute independently.
As shown in \Cref{fig:control_flow_separation}, the original VLIW program must execute each VLIW instruction sequentially, creating false dependencies between operations on different \saname{}s even though they do not have any true data dependencies.
As the compiler explicitly specifies how many \saname{}s are being used, the allocated \saname{}s cannot be changed at runtime 
unless the DNN program is recompiled.
For example, if the compiler generates \texttt{push}/\texttt{pop} 
operations for two \saname{}s, these operations cannot be time-multiplexed on a single \saname{}, since this will corrupt the intermediate states in the \saname{}.
Hence, if only one \saname{} is available, this DNN program cannot run 
until at least two \saname{}s are available (\Cref{{fig:isa_limitation}} left).
It also cannot utilize more than two \saname{}s, even if more than two are available (\Cref{{fig:isa_limitation}} right), because the \texttt{push}/\texttt{pop} operations for 
one \saname{} share the intermediate data in this \saname{}. 

To address this problem and enable dynamic \saname{} scheduling, one may consider switching from VLIW to another ISA (e.g., RISC) or employing superscalar out-of-order (OoO) execution (similar to a CPU core). However, they still lack the support for dynamic \saname{} scheduling since the compiler still needs to specify which \saname{} is the target of a \texttt{push}/\texttt{pop} instruction statically.
To remove such a constraint, we need to offer the flexibility for the NPU program to determine the target ME at runtime. Therefore, we need to rethink the contract between the compiler and the NPU hardware by extending the ISA.

\begin{figure}[t]
    \centering
    \includegraphics[width=\linewidth]{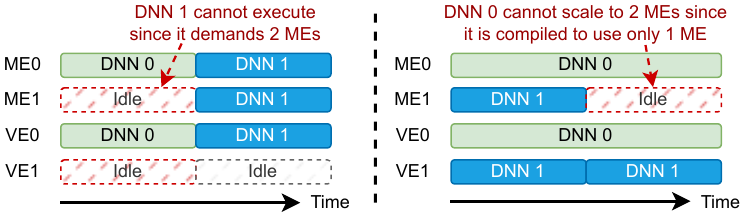}
    \caption{The current VLIW-style ISA causes NPU underutilization, as it cannot exploit available \saname{s} at runtime.}
    \label{fig:isa_limitation}
\end{figure}

\vspace{0.2em}
\noindent
\textbf{Architectural support for parallelizing \saname{}/\vuname{} operations.}
Our key observation is that the execution of different \saname{s} and \vuname{s} in a tensor operator 
is usually \textit{separable}. 
Specifically, most DNN operators, such as matrix multiplication (MatMul) and convolution, are partitioned by DNN compilers~\cite{tvm,roller:osdi2022} into multiple tiles that can be computed independently.
As shown in \Cref{fig:control_flow_separation}, the original program computes a MatMul tile and directly applies a 
ReLU function to the results using 2 \saname{}s and 2 \vuname{}s. However, the 
instructions executed on the first \saname{}/\vuname{} (colored blue) have no dependencies with the instructions 
on the second \saname{}/\vuname{} (colored green). The two instruction groups can be separated 
and independently executed. 

\section{Design and Implementation}
\label{sec:design}


We design \pname{} to achieve the following objectives:
\begin{itemize}[leftmargin=*]
    \item \textbf{Allocation flexibility:} As DNN workloads have different resource and SLO 
    requirements, we need to provide the flexibility for users to customize their NPU hardware.
    \item \textbf{NPU utilization:} Since an individual ML inference workload underutilizes NPU cores ($\S$\ref{sec:npu_util_study}), we need to enable fine-grained NPU virtualization for improved NPU utilization. 
    \item \textbf{Performance isolation:} As we collocate DNN workloads on the same NPU core, we must provide 
    performance isolation. 
\end{itemize}


We first present a new vNPU abstraction for NPU virtualization ($\S$\ref{sec:vnpu_abstraction}). 
Based on this, we enable flexible vNPU resource allocation ($\S$\ref{sec:vnpu_allocation}) and vNPU-to-pNPU mappings ($\S$\ref{sec:vnpu_mapping_hw}). 
We extend VLIW-style ISA ($\S$\ref{sec:isa_extension}) and NPU architecture ($\S$\ref{sec:arch_for_neuISA}) for enabling fine-grained 
resource scheduling for vNPUs.

\subsection{vNPU: The New Abstraction for NPU Virtualization}\label{sec:vnpu_abstraction}


We design the vNPU abstraction with the goals of 
(1) allocating NPU hardware resource 
to a vNPU instance on demand; 
(2) hiding the complexity from the ML programs with minimal changes to the guest software stack for compatibility.

\begin{figure}[t]
    \centering
    \includegraphics[width=0.9\linewidth]{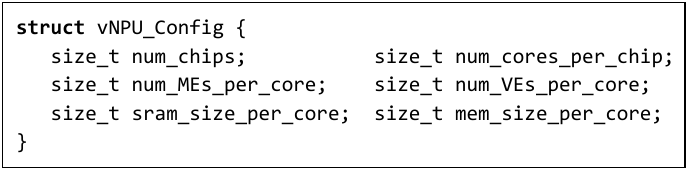}
	\caption{vNPU configuration.}
    \label{fig:vnpu_configuration}
\end{figure}

\vspace{0.2em}
\noindent
\textbf{vNPU abstraction.}
A vNPU instance reflects the hierarchy of a physical NPU board.
\Cref{fig:vnpu_configuration} shows the configurable parameters of a vNPU.
Each vNPU is exposed to the VM as a PCIe device. 
The guest NPU driver can query the hierarchy of the vNPU, such as the number of chips, cores per chip, HBM size, and others. 
The maximum vNPU size is capped by the physical NPU size. If a guest VM
requires more resources than is available on a physical NPU
board, \pname{} can allocate multiple vNPU instances to it.
The guest ML framework can handle the data distribution across multiple vNPU cores in the same way as that on physical NPUs.
Take Google TPU for example,
TensorFlow already handles 
data parallelism across physical NPUs. It can work in the same way with vNPUs. 


\begin{figure}[t]
    \centering
    \includegraphics[width=\linewidth]{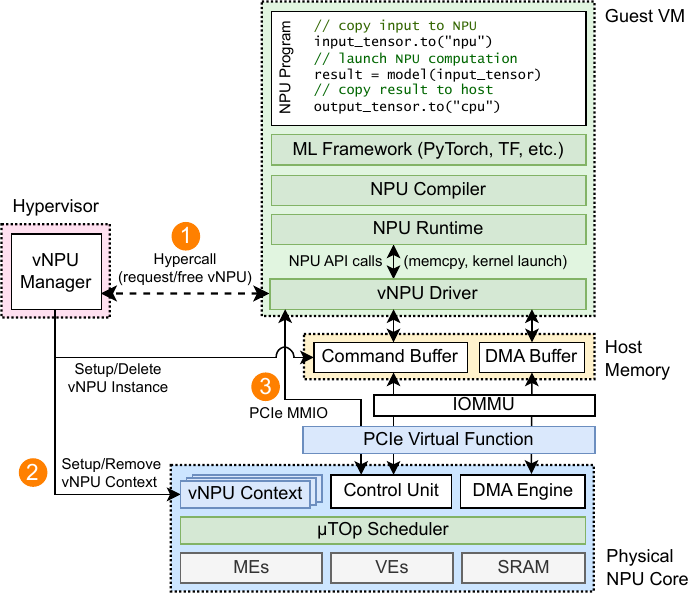}
    \caption{System architecture of \pname{}.}
    \label{fig:vnpu_interface}
\end{figure}

\vspace{0.2em}
\noindent
\textbf{vNPU lifecycle.}
To create a vNPU instance, a user can specify the vNPU configuration following the pay-as-you-go model~\cite{pay-as-you-go}.
Cloud providers can define various default configurations (e.g., small/medium/large vNPU cores as 
having 1/4/8 \saname{}s/\vuname{}s).
\pname{} can also learn an optimized vNPU configuration
for a DNN workload with ML compilers ($\S$\ref{sec:vnpu_allocation}).
As shown in \Cref{fig:vnpu_interface}, upon vNPU initialization, the guest driver sends a request to the hypervisor through a para-virtualized interface (\S\ref{sec:sys_virt}) (\circleo{1}).
The vNPU manager maps the vNPU instance to NPU hardware resources ($\S$\ref{sec:vnpu_mapping_hw}). 
Then, it initializes the vNPU context in the physical NPU device and creates the MMIO mappings for the guest VM to access the vNPU (\circleo{2}).
During execution, the application issues commands such as memcpy and compute offloading through the command buffer.
The NPU hardware directly fetches the commands from the host memory without the hypervisor intervention. 
It also has DMA access to the DMA buffer in the guest memory space via the IOMMU. 
The DNN program on the NPU executes asynchronously from the CPU program, and the NPU hardware schedules vNPUs ($\S$\ref{sec:arch_for_neuISA}) independently of existing OS/hypervisor schedulers.
The guest VM waits for the completion interrupt or 
actively polls the memory-mapped control registers for the current status of the vNPU (\circleo{3}).
After execution, the user can free the vNPU. 
\subsection{vNPU Allocation and Deallocation}\label{sec:vnpu_allocation}
Following the popular pay-as-you-go model~\cite{pay-as-you-go}, cloud platforms allow users to specify the vNPU configuration on demand. 
However, as ML inference workloads have diverse \saname{}/\vuname{} demands (see $\S$\ref{sec:npu_util_study}), 
specifying the number of \saname{}s/\vuname{}s can be challenging for users who are not NPU experts. Thus, 
we allow them to specify the total number of execution units (EUs), which is directly related to the cost of running the vNPU instance.
\pname{} provides the vNPU allocator, a compile-time tool to improve the performance per cost of vNPUs by identifying an optimized \saname{}/\vuname{} ratio for the user workload.


\vspace{0.2em}
\noindent\textbf{\saname{}/\vuname{} allocation.}
The \saname{}/\vuname{} demands of a ML workload can be reflected by how it runs on one \saname{} and one \vuname{}.  
We denote the $\frac {\textbf{\saname{} active runtime}} {\textbf{NPU total runtime}}$ as \emph{\textbf{m}}, and that of \vuname{} as \emph{\textbf{v}}.
These numbers can be obtained via profiling at the compilation stage. 
Based on our study in {$\S$\ref{sec:npu_util_study}}, for most DNN models, at least one of \saname{}/\vuname{} is active during the execution of an NPU core. Thus, the time portion where only \saname{} is active is $1-v$, that of only \vuname{} is $1-m$, and that of concurrent \saname{}/\vuname{} execution is $m+v-1$. 
With Amdahl's Law, the normalized execution time on $\boldsymbol{n_m}$ \saname{}s and $\boldsymbol{n_v}$ \vuname{}s is

\vspace{-1ex}
\begin{equation}
    \label{eq:norm_time}
    \vspace{-0.5ex}
    \boldsymbol{T} = \frac {1-v} {n_m} + \frac {1-m} {n_v} + \frac {m+v-1} {\min(n_m,n_v)}.
\end{equation}

\noindent
where the concurrent part is bottlenecked by the minority type of EU.
Let $n_m+n_v$ be the hypothetical speedup regardless of EU types, which means an EU can execute both \saname{} and \vuname{} operators.
Compared to real cases where each EU must respect data dependencies and operator types, the hypothetical speedup assumes all $n_m+n_v$ EUs are 100\% utilized.
Thus, the hypothetical execution time on $\boldsymbol{n_m}$ \saname{}s and $\boldsymbol{n_v}$ \vuname{}s is $\boldsymbol{T_h} = \frac {m+v} {n_m + n_v}$,
and the total EU utilization can be quantified as the ratio between hypothetical and estimated execution times:
\begin{equation}\label{eq:total_EU_util}
    \boldsymbol{U} = \frac {\boldsymbol{T_h}} {\boldsymbol{T}} = \frac{m+v} {(n_m+n_v)(\frac{1-v}{n_m}+\frac{1-m}{n_v}+\frac{m+v-1}{\min(n_m,n_v)})}.
\end{equation}

To isolate the impact of total \saname{} and \vuname{} quantity, we simplify the function by letting $\boldsymbol{k}=n_m/n_v$ be the ratio between 
the numbers of \saname{}s and \vuname{}s. Without loss of generality, we assume $n_v \geq n_m$, which means $k \leq 1$.
Then, we can simplify \Cref{eq:total_EU_util} with mathematical tools~\cite{wolfram_alpha}:

\begin{equation}
    \boldsymbol{U} =  \frac {(m+v)k} {(1-m)k^2+k+m} \quad (k \leq 1).
\end{equation}
To find the value of $k$ that maximizes $U$, we compute the value of $k$ where $\frac{\mathrm{d}U}{\mathrm{d}k}=0$.
This gives $k=\sqrt{m/(1-m)}$ for $m < 0.5$. If $m \geq 0.5$, $U$ will be monotonic, so $k=1$ maximizes $U$.
Similarly, for the case when $n_m \geq n_v$, we derive $k=\sqrt{(1-v)/v}$ for $v<0.5$ and $k=1$ for $v \geq 0.5$.
Consequently, we have
\begin{equation}\label{eq:vnpu_allocation}
    k = \frac{n_m}{n_v} = \begin{cases}
        \sqrt{m/(1-m)}, &m<0.5, \\
        \sqrt{(1-v)/v}, &v<0.5, \\
        1, &m\ge0.5 \text{ and } v\ge0.5.
    \end{cases}
\end{equation}
The case when both $m<0.5$ and $n<0.5$ does not exist since at least one of \saname{}/\vuname{} will be active ($m+n \geq 1$).
When $m < 0.5$, for workloads with \saname{} active time ratio $m$, we allocate  
$\sqrt{m/(1-m)}$ times more \saname{}s than \vuname{}s. When $v < 0.5$, for workloads with \vuname{} active 
time ratio $v$, we approximate the allocated \saname{}/\vuname{} quantity ratio to $\sqrt{(1-v)/v}$. If $m>0.5$ and $v>0.5$, 
we allocate the same number of \saname{}s and \vuname{}s.
Note that each vNPU will have at least one \saname{} and one \vuname{}.

\vspace{0.2em}
\noindent\textbf{Memory allocation.}
Users can use the compiler to estimate the total HBM capacity needed by a DNN workload. By default, the SRAM capacity is allocated proportionally to the number of allocated \saname{s}, as 
more \saname{s} usually indicate larger tile sizes. Based on our study 
in $\S$\ref{sec:npu_util_study}, for many common ML inference services, the HBM bandwidth is less of a concern. 
Thus, \pname{} allows fair sharing of HBM bandwidth by default. For large models that demand large HBM capacity and bandwidth, the vNPU abstraction offers the flexibility for end users to allocate the demanded resources.
The user may also leverage existing tensor swapping techniques to support large DNN workloads with limited memory capacity~\cite{swapadvisor:asplos20, g10:micro23}.
After vNPU allocation, ML compilers will compile the DNN program with the allocated resources. The compiler ensures the DNN program does not exceed the allocated SRAM and HBM. We will discuss how \pname{} handles compilation for different numbers of \saname{}s/\vuname{}s in $\S$\ref{sec:isa_extension}.

\begin{figure}[t]
    \centering
    \begin{subfigure}{0.49\linewidth}
        \includegraphics[width=0.85\linewidth]{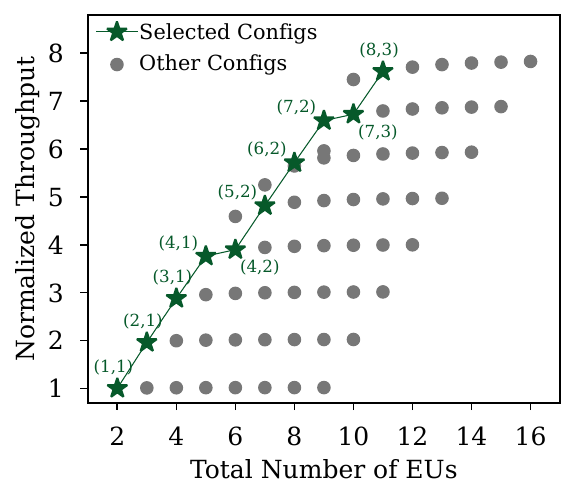}
        \caption{BERT (batch size 32).}
    \end{subfigure}
    \hfill
    \begin{subfigure}{0.49\linewidth}
        \includegraphics[width=0.85\linewidth]{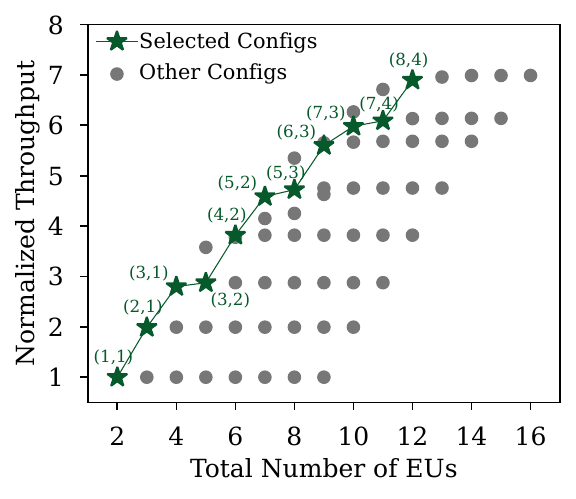}
        \caption{ResNet (batch size 32).}
    \end{subfigure}
    \newline
    \begin{subfigure}{0.49\linewidth}
        \includegraphics[width=0.85\linewidth]{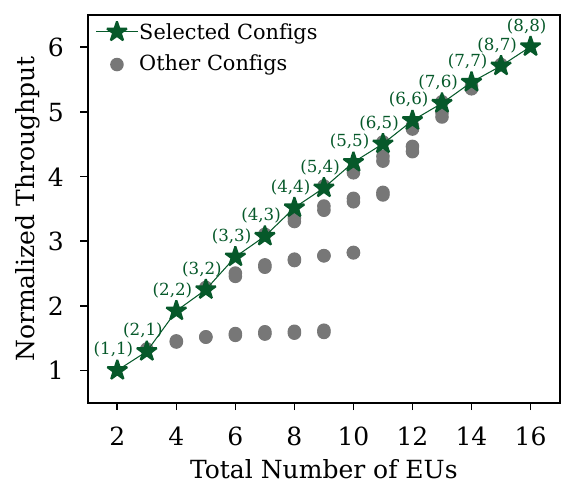}
        \caption{EfficientNet (batch size 32).}
    \end{subfigure}
    \hfill
    \begin{subfigure}{0.49\linewidth}
        \includegraphics[width=0.85\linewidth]{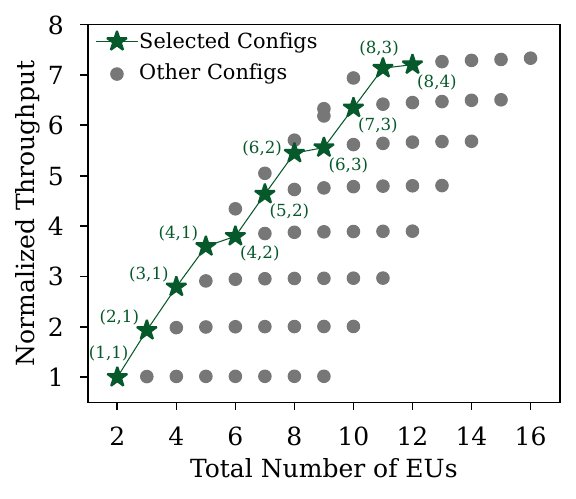}
        \caption{ShapeMask (batch size 8).}
    \end{subfigure}
    \caption{vNPU allocation results for representative DNN models as we scale up the available EUs on an NPU core from 1 {\saname{}} and 1 {\vuname{}} to 8 {\saname{}}s and 8 {\vuname{}}s. Each data point is a vNPU configuration. The label ($m$, $v$) means $m$ \saname{}s and $v$ \vuname{}s.}
    \label{fig:eval_vnpu_allocation}
\end{figure}

\vspace{0.2em}
\noindent\textbf{Cost-effectiveness analysis.}
We evaluate our allocation algorithm in \Cref{fig:eval_vnpu_allocation}. For vNPUs with no more than 4 \saname{}s and 2 \vuname{}s, we use a real TPUv4 to test the throughput. For others, we use a production-level NPU simulator (see \S\ref{sec:impl}).
In most cases, our algorithm selects a configuration with better performance than others for the same number of EUs. 
Though a sub-optimal configuration may be selected, it still achieves similar performance as the optimal one.
The \saname{}/\vuname{} harvesting ({$\S$\ref{sec:arch_for_neuISA}}) also tolerates some allocation inaccuracies by opportunistically utilizing more EUs.

\vspace{0.2em}
\noindent\textbf{vNPU deallocation.}
Upon vNPU deallocation, the vNPU 
manager will send a command associated with the vNPU ID to the corresponding NPU board to clean up the vNPU context, as well as remove the DMA setup for this vNPU.



\subsection{vNPU Mapping}\label{sec:vnpu_mapping_hw}



The vNPU manager attempts to balance the number of allocated EUs and the size of allocated memory.  
This minimizes the chance that all EUs on one core are allocated but a large portion of its memory is not allocated, or vice versa.
Thus, vNPUs with many EUs and small memory will be collocated with vNPUs with few EUs and large memory. \pname{} uses a 
greedy algorithm for this by default. 

\vspace{0.2em}
\noindent
\textbf{vNPU mapping schemes.}
\pname{} provides the flexibility for cloud platforms to enable both \textit{hardware-isolated} (spatial-isolated) mapping 
and \textit{software-isolated} (temporal-sharing) mapping. With hardware-isolated mapping, a vNPU is mapped to dedicated EUs and SRAM, and the allocated hardware is not shared with other vNPUs.
With software-isolated mapping, multiple vNPUs can temporally share the same EUs.
\pname{} uses priority-based scheduling for fair sharing
and performs context switches between vNPUs (see $\S$\ref{sec:arch_for_neuISA}).

\vspace{0.2em}
\noindent
\textbf{vNPU mapping policies.}
{\pname{}} decides which vNPUs can be mapped onto the same physical NPU (pNPU) as follows.
With hardware-isolated mapping, {\pname{}} collocates a set of vNPUs as long as the total resource requirement (e.g., number of MEs/VEs, HBM capacity) does not exceed the pNPU.
With software-isolated mapping, {\pname{}} aims to load-balance the pNPUs while allowing oversubscription. {\pname{}} tracks the total resource requirement of assigned vNPUs on each pNPU, and assigns a new vNPU to the pNPU that suffers the least resource requirement. {\pname{}} can support other collocation policies~\cite{bubbleup:micro11,v10:isca23,prophet:asplos2017} as well.
At scale, {\pname{}} can be integrated with a cluster-wise VM/container orchestration framework such as KubeVirt/Kubernetes~\cite{kubevirt} to decide which VM should be placed on what machine. Developing advanced vNPU/VM collocation policies is orthogonal to our work.

\vspace{0.2em}
\noindent
\textbf{vNPU security isolation.}
\pname{} enforces memory address space isolation among collocated vNPUs with the conventional memory segmentation scheme~\cite{v10:isca23,memory_segmentation:wikipedia} for both HBM and SRAM.
\pname{} divides the SRAM and HBM into fixed-sized segments and maps each segment to the virtual address space of a vNPU.
For the NPU core in \Cref{tab:simulator_config}, an SRAM/HBM segment is 2MB/1GB.
There is no external fragmentation since the segment size is fixed.
The address translation is performed by adding the segment offset to the starting address of the physical segment, which incurs negligible overhead.
A page fault will be triggered when an invalid access happens. 
This is sufficient since ML frameworks like TensorFlow typically request a contiguous chunk of memory for the entire lifetime of an ML inference service and have their own memory management mechanism.
To isolate the vNPU instances as they communicate with the host,
\pname{} uses IOMMU to enforce DMA remapping ($\S$\ref{sec:sys_virt}).
We leave side-channel mitigation to future work.

\begin{figure}[t]
    \centering
    \includegraphics[width=\linewidth]{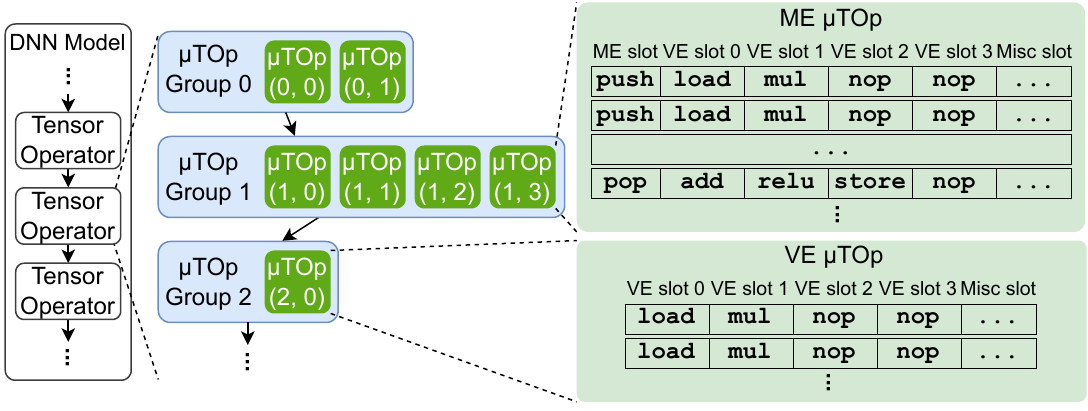}
    \caption{NeuISA programming model.}
    \label{fig:isa_programming_model}
\end{figure}

\begin{figure*}[t]
    \centering
    \includegraphics[width=\linewidth]{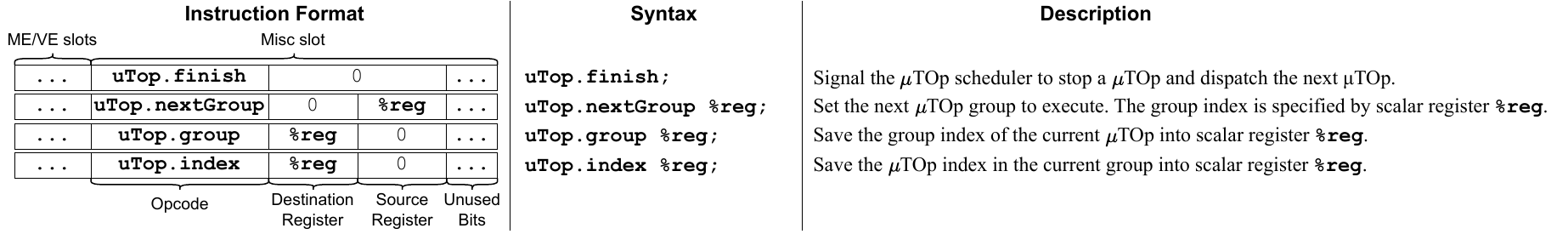}
    \caption{Definitions of \utop{} control instructions. Scalar register zero (\texttt{\%r0}) is read-only and always has a value of 0.}
    \label{fig:isa_control_insts}
\end{figure*}

\subsection{ISA Extension for NPU Virtualization}
\label{sec:isa_extension}


To support dynamic \saname{}/\vuname{} scheduling ($\S$\ref{subsec:challenges}), 
we develop NeuISA, in which when 
the ML compiler maps a tensor operator onto \saname{}s, it generates ``sub-tasks'' 
for each \saname{}, so the hardware can decide which ``sub-task'' can be 
executed at runtime based on the availability of \saname{}s. NeuISA is still  
expressive for compilers to exploit the instruction-level parallelism between \saname{}s and \vuname{}s, 
and preserve the flexibility of supporting fused operators and complex control-flow structures like branches 
and nested loops in VLIW-style ISAs.

\vspace{0.2em}
\noindent\textbf{Separating \saname{} control flow with \bfutop{}s.}
NeuISA decouples the execution of independent \saname{}s in a tensor operator by separating the control flow of each \saname{} and \vuname{} into independent instruction sequences (see~\Cref{fig:control_flow_separation}), 
called \textit{micro-Tensor Operators (\utop{}s)}.
To minimize changes to the existing VLIW compiler and hardware, the instruction format inside a \utop{} resembles the original 
VLIW ISA: an instruction contains multiple slots, and each slot encodes an 
operation (such as a \texttt{push}/\texttt{pop} operation in an \saname{} slot and an ALU operation in a \vuname{} slot). 
However, the number of \saname{} slots in a \isa instruction differs from that of a traditional NPU ISA.

\vspace{0.2em}
\noindent\textbf{\bfutop{} types.}
As shown in \Cref{fig:isa_programming_model}, for a physical NPU core with $n_x$ \saname{}s and $n_y$ \vuname{}s, NeuISA defines two types of \utop{}s: 
(1) An \textit{\saname{} \utop{}} contains instructions with one \saname{} slot and $n_y$ \vuname{} slots. 
An \saname{} \utop{} will only use one \saname{} during execution, which enforces that each \saname{} \utop{} only contains the control flow of one \saname{}.
To execute an operator on multiple \saname{}s, 
the compiler generates multiple \saname{} \utop{}s.
At runtime, the hardware dynamically adjusts the number of \saname{}s 
assigned to this operator by deciding how many \saname{} \utop{}s are being executed.
The \vuname{} slots in an \saname{} \utop{} enable instruction-level parallelism between \saname{}s and \vuname{}s.
\vuname{} slots are necessary because the \vuname{} needs to aggregate the outputs of the systolic array. They also enable operator fusions such as MatMul+ReLU (see \Cref{fig:control_flow_separation}).
(2) A \textit{\vuname{} \utop{}} contains instructions with no \saname{} slot and $n_y$ \vuname{} slots,
which performs vector operations that do not involve \saname{} computation. 
The $n_y$ \vuname{} slots allow a \vuname{} \utop{} to utilize all the \vuname{}s.
Having multiple \vuname{} slots in an instruction does not increase the hardware complexity since the original VLIW NPU architecture already supports this.

\vspace{0.2em}
\noindent\textbf{Supporting fused operators with \bfutop{} groups.}
The \utop{}s can efficiently support basic tensor operators, such as tiled matrix multiplication with each \utop{} 
computing a different tile. However, ML compilers may generate fused operators that cannot be handled by \utop{}s alone, e.g., a matrix multiplication may be executed with $n_x$ \saname{} \utop{}s, 
while the succeeding fused normalization operator only needs a \vuname{} \utop{}.

To support a fused operator, NeuISA organizes the \utop{}s into a sequence of \textit{\utop{} groups} to express the 
dependencies between \utop{}s, as shown in~\Cref{fig:isa_programming_model}. Each group contains up to $n_x$ \saname{} \utop{}s, 
allowing the operator to utilize all the allocated \saname{}s, and up to one \vuname{} \utop{}, as one \vuname{} \utop{} 
already contains $n_y$ \vuname{} slots to utilize all the \vuname{}s. All \utop{}s in one \utop{} group may execute concurrently, 
but each group must execute sequentially to preserve data dependency.
As an example, a fused operator may contain one \utop{} group doing a MatMul+ReLU with 
multiple \saname{} \utop{}s, followed by a \utop{} group doing normalization with a single \vuname{} \utop{}.


\vspace{0.2em}
\noindent
\textbf{NeuISA control flow.}
As NeuISA inherits the VLIW semantic inside each \utop{}, it intrinsically supports conditional branches and loops inside a \utop{}.
It is also desirable to have branches across \utop{} groups. For example, an operator contains a nested loop in which the inner-most loop 
is a matrix multiplication that can be mapped to a \utop{} group. In this case, we need to support loops across multiple \utop{} groups.

\begin{figure}[t]
    \centering
    \includegraphics[width=\linewidth]{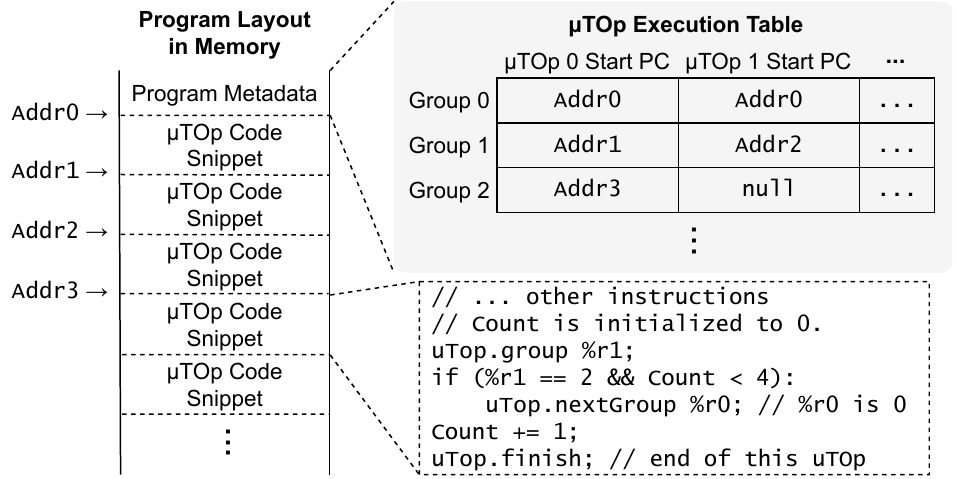}
    \caption{\isa{} program structure.}
    \label{fig:isa_prog_structure}
\end{figure}

NeuISA defines special control instructions that can be invoked in each \utop{} (see \Cref{fig:isa_control_insts}).
The \texttt{uTop.nextGroup} instruction can be used to specify the target \utop{} group that should be executed next. It may be executed by more than one \utop{}s in the same group as long as they specify the same target group index.
Otherwise, an exception will be raised.
\Cref{fig:isa_prog_structure} shows a loop structure example.
The loop counter \texttt{Count} is stored in the on-chip SRAM. 
The loop body contains \utop{} group 0--2. 
In group 2, \texttt{Count} is incremented and examined at the end of a \utop{}. If this is not the last loop iteration, \texttt{uTop.nextGroup} is executed to loop back to group 0.

\vspace{0.2em}
\noindent
\textbf{NeuISA program structure.}
A NeuISA binary contains \textit{\utop{} code snippets}, as shown in \Cref{fig:isa_prog_structure}, which are the assemblies for \utop{}s.
The \utop{} groups are encoded by a \textit{\utop{} execution table}. 
Each row defines a \utop{} group. Each cell is the start address of a \utop{} code snippet.
NeuISA provides control instructions to retrieve the group index and \utop{} index of the current \utop{} (see~\Cref{fig:isa_control_insts}).
The size of each row in the \utop{} execution table depends on the number of \saname{}s/\vuname{}s on the physical NPU core. For a physical core with $n_x$ \saname{}s, each row has $n_x$ \saname{} \utop{} entries and one \vuname{} \utop{} entry. An entry will be \texttt{null} if the \utop{} does not exist in the group.

A DNN program is executed by the NPU core following the \utop{} execution table. By default, \utop{} group $i+1$ will be executed after group $i$ (starting from group 0), unless \texttt{uTop.nextGroup} specifies another group index. The \utop{}s in the same group can execute in any order.
Each \utop{} executes a snippet of VLIW instructions.

\begin{figure}[t]
    \centering
    \includegraphics[width=\linewidth]{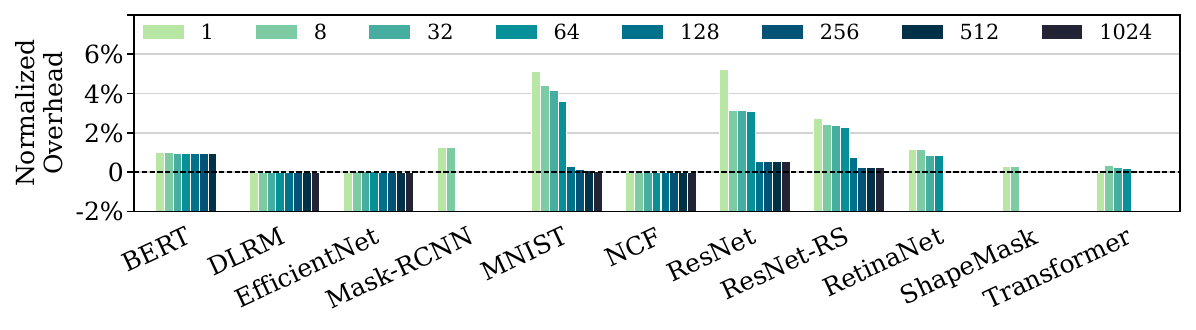}
    \caption{Performance overhead of NeuISA over the traditional VLIW-style ISA for various DNN workloads.}
    \label{fig:neuISA_overhead}
\end{figure}

\vspace{0.2em}
\noindent
\textbf{Compiler support for NeuISA.}
\isa{} allows a DNN program to utilize different numbers of \saname{}s/\vuname{}s at runtime without recompilation, regardless of the allocated vNPU size at compilation time.
This is supported with minimal compiler changes. 
For a physical NPU core with $n_x$ \saname{}s and $n_y$ \vuname{}s, we first employ existing compiler techniques~\cite{roller:osdi2022} 
to partition each operator into up to $n_x$ \utop{}s, which allows the DNN program to utilize all \saname{}s on the NPU core. Next, we employ the existing compiler backend such as XLA~\cite{xla} 
to compile each \utop{} independently assuming a fictional NPU with one \saname{} and $n_y$ \vuname{}s.
Finally, we extract the dependencies between \utop{}s from the DNN execution graph, and append \isa control flow instructions at the end of \utop{}s when necessary.

\vspace{0.2em}
\noindent
\textbf{\isa Overhead.}
\isa incurs negligible performance overhead (less than 1\% on average) for most DNN workloads (see \Cref{fig:neuISA_overhead}).
The major overhead occurs when a matrix multiplication is partitioned on the reduction dimension to utilize all \saname{}s.
In this case, \isa prevents instruction-level pipelining between \saname{} computation and summing the \saname{} outputs on the \vuname{}s, as the summation must be done in a separate \vuname{} \utop{} after the \saname{} \utop{}s.
The overhead is smaller for larger batch sizes, as the compiler will partition other dimensions (e.g., the batch dimension) if they are large enough.
While \isa{} may inflate the code size by having multiple multiple \vuname{} slots in a \utop{}, this is less of a concern in practice since \isa minimizes code inflation by sharing the same code snippet among \utop{}s. The on-chip instruction memory is large enough to avoid stalling the pipeline.



\begin{figure}[t]
    \centering
    \includegraphics[width=0.7\linewidth]{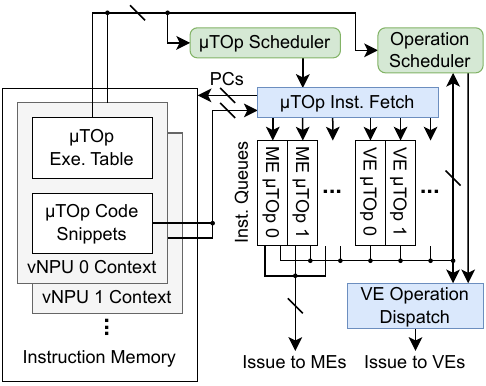}
    \caption{NPU core pipeline frontend for NeuISA.}
    \label{fig:isa_pipeline}
\end{figure}

\subsection{Architectural Support for NeuISA}
\label{sec:arch_for_neuISA}
The \utop{} design enables dynamic operator scheduling.
It allows a vNPU to harvest unused \saname{}/\vuname{}s from other collocated vNPUs in the same physical NPU core at runtime.

\vspace{0.2em}
\noindent
\textbf{Hardware scheduler for \isa.}
\Cref{fig:isa_pipeline} shows the pipeline design for fetching and scheduling \utop{}s.
The NPU core maintains the contexts of multiple vNPUs, including the PC pointers to the program and the vNPU configurations.
Each time a new \utop{} is ready or an existing \utop{} finishes, the \textit{\utop{} scheduler} selects the \utop{}s to be executed next. 
For each vNPU, the \utop{} scheduler retrieves the number of allocated \saname{}s and the number of ready \saname{} \utop{}s from the vNPU context.
It selects a set of ready \utop{}s, 
and fetch their instructions to the instruction queues.

Next, the \textit{operation scheduler} selects which operations from the instruction queues will be executed at every cycle.
The \saname{} operations
from the \saname{} \utop{} instruction queues
are directly issued to the corresponding \saname{}s.
For the \vuname{} operations, the scheduler selects which operations to issue from all \vuname{} \utop{} instruction queues.
To reclaim a harvested \saname{}, \pname{} performs a context switch to preempt the harvesting \utop{}.
Upon a context switch, the register file and the intermediate data in the \saname{}s are saved to SRAM, which incurs negligible overhead compared to the length of an operator.
The number of instruction queues should be large enough to support simultaneous execution of all \saname{}s/\vuname{}s.
For an NPU core with $n_x$ \saname{}s and $n_y$ \vuname{}s, there are $n_x$ \saname{} \utop{} instruction queues and $n_y$ \vuname{} \utop{} instruction queues.


\vspace{0.2em}
\noindent
\textbf{\bfutop{} scheduling policy.}
The \utop{} scheduler can be configured in either spatial-isolated or temporal-sharing vNPU scheduling mode, as discussed in $\S$\ref{sec:vnpu_mapping_hw}.

With spatial-isolated mode, the scheduler aims to ensure performance isolation.
First, if a vNPU has $n_x$ \saname{}s and at least $n_x$ ready \saname{} \utop{}s, 
the scheduler will execute $n_x$ \saname{} \utop{}s to fully utilize all the allocated \saname{}s for this vNPU.
In this case, no \saname{}s will be harvested from this vNPU. If the allocated \saname{}s are already being harvested by \utop{}s from other vNPUs, these \utop{}s will be preempted to reclaim the harvested \saname{}s.
Second, to improve utilization, if the vNPU has more than $n_x$ ready \saname{} \utop{}s, and if another 
vNPU does not have enough \saname{} \utop{}s to utilize all its \saname{}s, the scheduler allows the 
unused \saname{s} to be harvested.  
A ready \vuname{} \utop{} is always executed, as it does not occupy any \saname{}s.

With temporal-sharing mode, as the NPU is oversubscribed, the scheduler maintains fair sharing with the best effort. It uses a priority-based preemptive policy similar to that in previous works~\cite{v10:isca23,prema:hpca20}.
It uses a performance counter to track the active cycles of each vNPU and balances the execution times of vNPUs based on their relative priorities.



\begin{figure}[t]
    \centering
    \includegraphics[width=\linewidth]{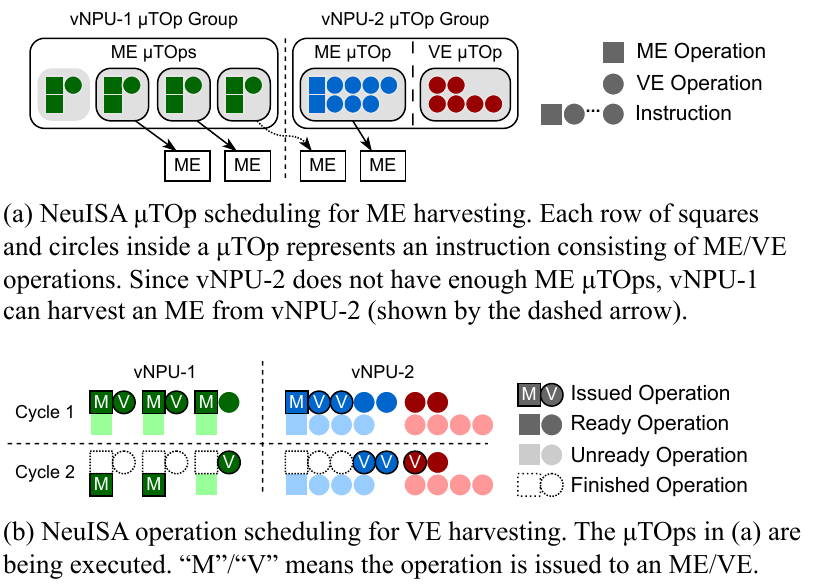}
    \caption{\saname{}/\vuname{} harvesting in \isa. For simplicity, we assume all operations finish in one cycle. In practice, an \saname{} operation takes longer time than a \vuname{} operation (see $\S$\ref{subsec:npuarch}).}
    \label{fig:isa_scheduling}
\end{figure}

\Cref{fig:isa_scheduling}(a) shows an example of two vNPUs collocated on an NPU with 4 \saname{}s 
and 4 \vuname{}s with spatial-isolated mapping. Each vNPU has 2 \saname{}s and 2 \vuname{}s.
Since vNPU-2 only has one \saname{} \utop{}, vNPU-1 can harvest an \saname{} from vNPU-2.


\vspace{0.2em}
\noindent
\textbf{Operation scheduling policy.}
The operation scheduler schedules \vuname{} operations using a similar policy as \saname{} \utop{} scheduling.
First, the scheduler determines the number of \vuname{}s assigned to each vNPU.
Then, among all the \vuname{} operations in each vNPU, the scheduler prioritizes those from \saname{} \utop{}s, which allows the occupied \saname{}s to be freed as soon as possible.

\Cref{fig:isa_scheduling}(b) shows an example of \vuname{} scheduling. In cycle 1, vNPU-1 has 3 ready \vuname{} 
operations and vNPU-2 has 6 ready ones. Each vNPU has 2 \vuname{}s, and all \vuname{}s are given to operations from \saname{} \utop{}s.
In cycle 2, vNPU-1 has one ready \vuname{} operation, so one of its \vuname{}s is harvested by vNPU-2. 
Since vNPU-2 gets 3 \vuname{}s and its \saname{} \utop{} cannot utilize all of them, 
the remaining \vuname{} is given to the \vuname{} \utop{}.

\subsection{System Support for NPU Virtualization}\label{sec:sys_virt}
\noindent
\textbf{OS hypervisor.}
\pname{} can work with OS hypervisors to provide system support for virtualizing NPUs. Take the KVM hypervisor as a case study, 
\pname{} leverages \texttt{vfio-mdev} to expose vNPUs to VMs as mediated PCIe devices~\cite{vfio-mdev}. 
The hypervisor only mediates the resource management functions that are not on the critical path, including the following hypercalls: (1) create a new vNPU, (2) change the configuration
of an existing vNPU, 
and (3) deallocate a vNPU.
The hypercalls are routed to the vNPU manager, which is implemented as a host kernel module. The vNPU manager tracks the allocated and free resources (e.g., \saname{}s/\vuname{}s, SRAM, HBM) of all physical NPUs on the host machine and implements the vNPU mapping policies (\S\ref{sec:vnpu_mapping_hw}).
Once a vNPU is set up, the VM can bypass the hypervisor and directly talk to the NPU device.
\pname{} uses SR-IOV~\cite{sr-iov} to expose each vNPU as a PCIe virtual function to the VM via PCIe-passthrough. The IOMMU performs DMA and interrupt remapping for the vNPUs.

\vspace{0.2em}
\noindent
\textbf{Guest VM software.}
\pname{} requires minimal changes to the guest VM software stack. 
First, the user source code remains unchanged. 
Typically, user codes are programmed with ML frameworks like PyTorch or TensorFlow~\cite{PyTorch, tensorflow}.
Second, for ML frameworks, only the backend NPU compiler needs to be revised to support \isa{} (\S\ref{sec:isa_extension}). The ML framework has two parts: (1) The frontend converts the user code into a device-agnostic DNN dataflow graph and optionally partitions the graph onto multiple NPU cores. As our vNPU abstraction reflects the hierarchy of a physical NPU device, the frontend requires no changes.
(2) The backend compiles the DNN graph into NPU binary using \isa{}. 
Third, the NPU vendor will provide a para-virtualized vNPU driver, which is a common practice for virtualizing PCIe devices. The vNPU driver provides user APIs for vNPU management and issues hypercalls to realize these APIs. 
With PCIe passthrough, the vNPU driver can directly interact with the NPU device~\cite{vfio-mdev}. 


\subsection{\pname{} Implementation}\label{sec:impl}

We implement \pname{} with a production-level event-driven NPU simulator.
We obtain the operator execution traces for each DNN workload on real Google Cloud TPUs.
For each operator, the trace contains the \saname{}/\vuname{} time, HBM time, tensor shapes, and the tile sizes and tiling dimensions selected by the compiler.
We use the tiling information to generate \utop{}s and replay the generated {\utop{}} traces in our simulator.
We modify the frontend of the NPU simulator to implement the \utop{} scheduling and harvesting policy (see {\S\ref{sec:arch_for_neuISA}}).
The scheduler picks \utop{}s from multiple traces (each trace represents the DNN workload of a vNPU) and issues them to the backend, which simulates the execution of each \saname{}/\vuname{}, on-chip SRAM accesses, and DMA operations to the off-chip HBM at cycle level.
To model the penalty of \utop{} preemption (i.e., the context switch overhead of \saname{}s), we set the \saname{} preemption latency to 256 cycles based on the systolic array dimension (i.e., 128$\times$128),
including 128 cycles to pop the partial sums and 128 cycles to pop the weights of the preempted \utop{}.


We also prototype the hardware scheduler for \isa in Verilog and synthesize it using the FreePDK-15nm 
cell library~\cite{freepdk15}.
Since the DNN workloads have deterministic dataflow graphs, they do not require complex dependency tracking or speculation in hardware.
The hardware area overhead of \pname{} is only 0.04\% on a TPUv4 chip. 
The power overhead of this small extra area is negligible compared to that of the entire chip.

\section{Discussion}\label{sec:discussion}

\noindent
\textbf{Support for multi-chip inferences.}
Currently, \pname{} supports multi-chip inference with data parallelism by using multiple vNPU chips. As the first step of NPU virtualization, we focus on enabling fine-grained resource sharing on individual NPU chips. In future work, we will extend \pname{} by investigating how to virtualize inter-chip interconnects to support more complicated scenarios (e.g., model parallelism).

\vspace{0.2em}
\noindent
\textbf{Engineering efforts in developing \pname{}.}
While \pname{} is a full-stack NPU virtualization design, each component is developed and tested in a modular way, and we minimize changes to the existing system at each level to reduce the debugging and verification efforts. The compiler and hardware changes are minimized as \isa{} reuses the VLIW instruction format in each \utop{}.
The guest vNPU driver greatly resembles a native NPU driver thanks to PCIe pass-through, and the major change is the new hypercalls for vNPU management. The KVM hypervisor already provides extensibility for new PCIe devices, and we leverage this feature to integrate the vNPU manager.

\vspace{0.2em}
\noindent
\textbf{Inter-generational compatibility with \isa{}.}
\isa{} enables a DNN program to run on different numbers of \saname{}s/\vuname{}s without recompilation.
This greatly eases the effort to provide compatibility across generations of NPU hardware.
\isa{} could ease the future development efforts of ML frameworks to support new NPU hardware and enable more flexible and transparent ways to manage NPU resources.
\pname{} provides a general vNPU abstraction, which allows a vNPU to be mapped to different generations of NPU hardware.



\section{Evaluation}
\label{sec:eval}
Our evaluation shows that:
(1) \pname{} provides performance isolation with up to $4.6\times$ reduction in tail latency, while improving the ML service throughput by 1.4$\times$ over state-of-the-art NPU sharing approaches ($\S$\ref{sec:eval_isolation});
(2) It improves the NPU utilization by 1.2$\times$ ($\S$\ref{sec:eval_util});
(3) It scales as we change the number of \saname{}s/\vuname{}s ($\S$\ref{sec:eval_sens_me_ve});
(4) It benefits multi-tenant ML services with various HBM bandwidths ($\S$\ref{sec:eval_sens_hbm_bw}). 


\subsection{Experimental Setup}\label{sec:eval_setup}

\begin{table}[t]
    \centering
    \caption{NPU simulator configuration.}
    \begin{tabular}{|l|l|}
    \hline
        \# of \saname{}s/\vuname{}s & 4 \saname{}s \& 4 \vuname{}s \\\hline
        \saname{} dimension & $128 \times 128$ systolic array \\\hline
        \vuname{} ALU dimension & $128 \times 8$ FP32 operations/cycle \\\hline
        Frequency & 1050 MHz \\\hline
        On-chip SRAM & 128 MB \\\hline
        HBM Capacity \& Bandwidth & 64 GB, 1200 GB/s \\\hline
    \end{tabular}
    \label{tab:simulator_config}
\end{table}

We evaluate DNN workloads (see~\Cref{tab:workloads}) from MLPerf v2.1~\cite{reddi2019mlperf} and the official TPU reference models~\cite{tpu_supported_models}.
To test \pname{} under different workload combinations, we select workload pairs with low \saname{}/\vuname{} contention (\texttt{DLRM+SMask}, \texttt{DLRM+RtNt}, \texttt{NCF+RsNt}), medium contention (\texttt{ENet+SMask}, \texttt{BERT+ENet}, \texttt{ENet+MRCN}), and high contention (\texttt{ENet+TFMR}, \texttt{MNIST+RtNt}, \texttt{RNRS+RtNt}).
The batch size is 32 except for \texttt{MRCN} and \texttt{SMask} (batch size is 8 for them).
Each workload runs on a vNPU with 2 \saname{}s and 2 \vuname{}s.
We map two vNPUs to a physical NPU core with 4 \saname{}s and 4 \vuname{}s as listed in~\Cref{tab:simulator_config}. The SRAM and HBM capacity is evenly partitioned between the vNPUs.
To obtain steady-state performance, we run inference requests continuously for each workload until all collocated workloads have completed a certain number of requests.
We compare the following designs:
\begin{itemize}[leftmargin=*]

    \item \textbf{\pmt{}}~\cite{prema:hpca20}: temporal-sharing of the entire NPU core among multiple vNPUs. A preemptive fair scheduling mechanism is employed for performance isolation.
    
    \item \textbf{\ten{}}~\cite{v10:isca23}: temporal-sharing of all \saname{}s and \vuname{}s among the vNPUs, with a priority-based preemptive policy.
    The workload is compiled with the traditional VLIW-style ISA.
    If an \saname{} operator from one vNPU is running, only \vuname{}-only operators from collocated vNPUs can execute simultaneously.

    
    \item \textbf{\pname{}-NoHarvest (\nohar{})}: spatial-isolated vNPUs with dedicated \saname{}s/\vuname{}s without dynamic scheduling. This resembles existing static partitioning techniques such as NVIDIA Multi-instance GPU (MIG)~\cite{mig:nvidia}.
    
    \item \textbf{\pname{}}: spatial-isolated vNPUs with dynamic resource scheduling and harvesting enabled by NeuISA.

\end{itemize}

\begin{figure}[t]
    \centering
    \includegraphics[width=\linewidth]{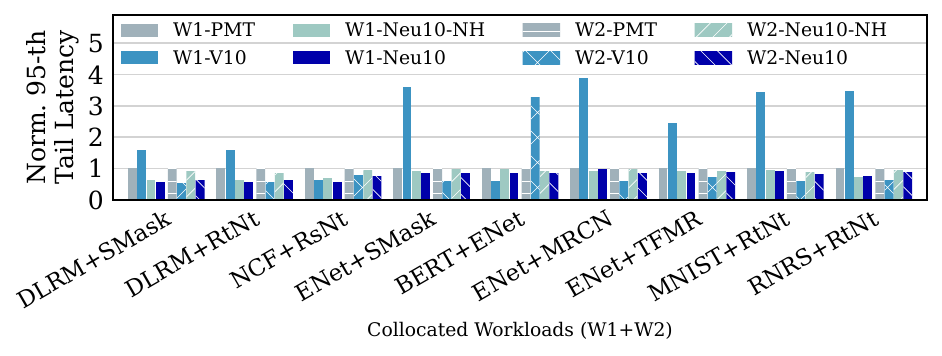}
    \caption{95\% Percentile latency of \pname{} (normalized to \pmt{}).}
    \label{fig:eval_tail_95}
\end{figure}

\begin{figure}[t]
    \centering
    \includegraphics[width=\linewidth]{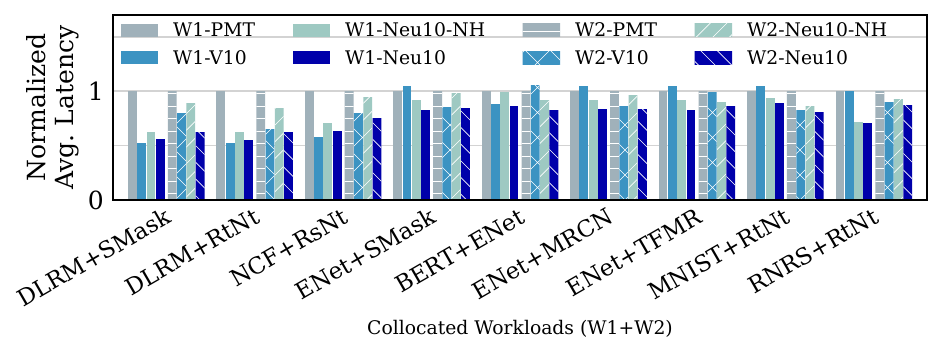}
    \caption{Average request latency of \pname{} (normalized to \pmt{}).}
    \label{fig:eval_avg_lat}
\end{figure}

\begin{figure}[!t]
    \centering
    \includegraphics[width=\linewidth]{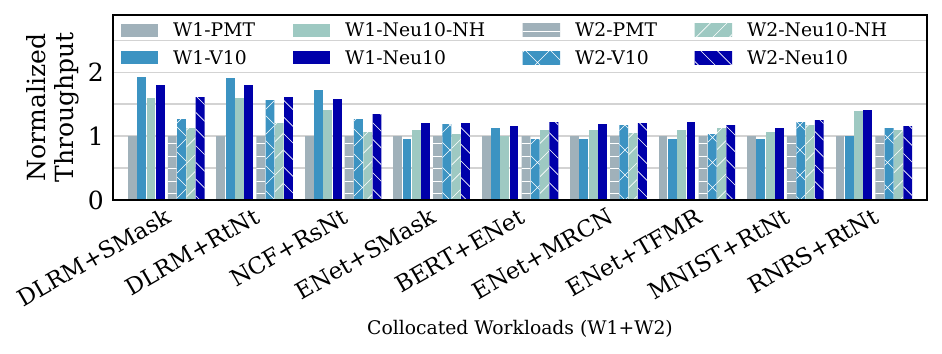}
    \caption{Throughput of \pname{} (normalized to \pmt{}).}
    \label{fig:eval_throughput}
\end{figure}

\begin{figure}[t]
    \centering
    \begin{subfigure}{\linewidth}
        \centering
        \includegraphics[width=\linewidth]{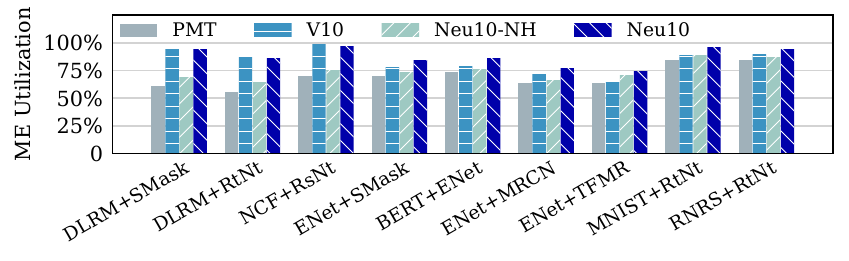}
        \vspace{-4ex}
        \caption{Total \saname{} utilization of the NPU core.}
    \end{subfigure}
    \vspace{0.1ex}
    \vfill
    \begin{subfigure}{\linewidth}
        \centering
        \includegraphics[width=\linewidth]{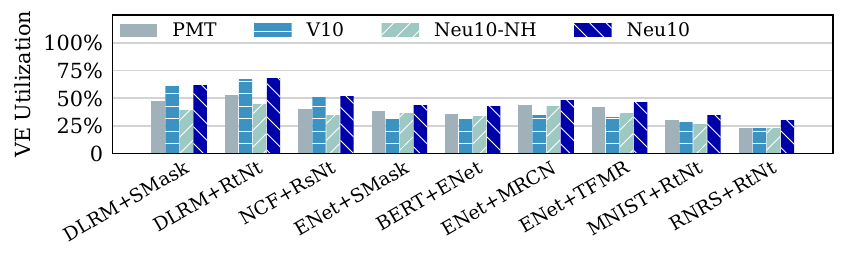}
        \vspace{-4ex}
        \caption{Total \vuname{} utilization of the NPU core.}
    \end{subfigure}
    \caption{Total utilization of \saname{}s and \vuname{}s.}
    \label{fig:eval_util}
\end{figure}


\subsection{Performance of \pname{}}\label{sec:eval_isolation}

\noindent\textbf{Tail Latency.}
\Cref{fig:eval_tail_95} shows that \pname{} improves the 95\% tail latency over \ten{} by up to $4.6\times$ ($1.56\times$ on average).
\ten{} primarily focuses on maximizing the utilization. Thus, even with an operator preemption mechanism,
it still fails to enforce performance isolation between vNPUs, due to complex inter-operator dependencies and imbalanced operator lengths.

In contrast, \pname{} ensures performance isolation between vNPUs while opportunistically improving their performance by harvesting. 
As \pname{} only harvests the underutilized compute units, the performance interference between vNPUs is minimized.
Hence, a harvested workload in \pname{} experiences negligible tail latency overhead compared to that in \nohar{}. In a few cases (e.g., \texttt{ENet+MRCN} and \texttt{RNRS+RtNt}), harvesting increases the burden on memory bandwidth, which may slightly impact the tail latency. However, \pname{} still achieves much better tail latency than \pmt{} and \ten{}.

\vspace{0.2em}
\noindent\textbf{Average Latency.}
\pname{} improves the average latency of inference requests by 1.33$\times$ over \pmt{} and 1.12$\times$ over \ten{} on average (\Cref{fig:eval_avg_lat}). 
While both \ten{} and \pname{} perform dynamic scheduling to utilize the NPU hardware, 
\pname{} greatly reduces \saname{} contentions with \utop{}-level scheduling.
\ten{} treats all \saname{}s on a physical NPU core as a whole unit, due to the VLIW ISA limitation. 
This causes false contentions on the \saname{}s when an operator cannot fully exploit the \saname{}s but still fully occupies them. 
In contrast, \pname{} eliminates such contention by assigning \saname{}s to different operators.

\begin{figure}[t]
    \centering
    \includegraphics[width=\linewidth]{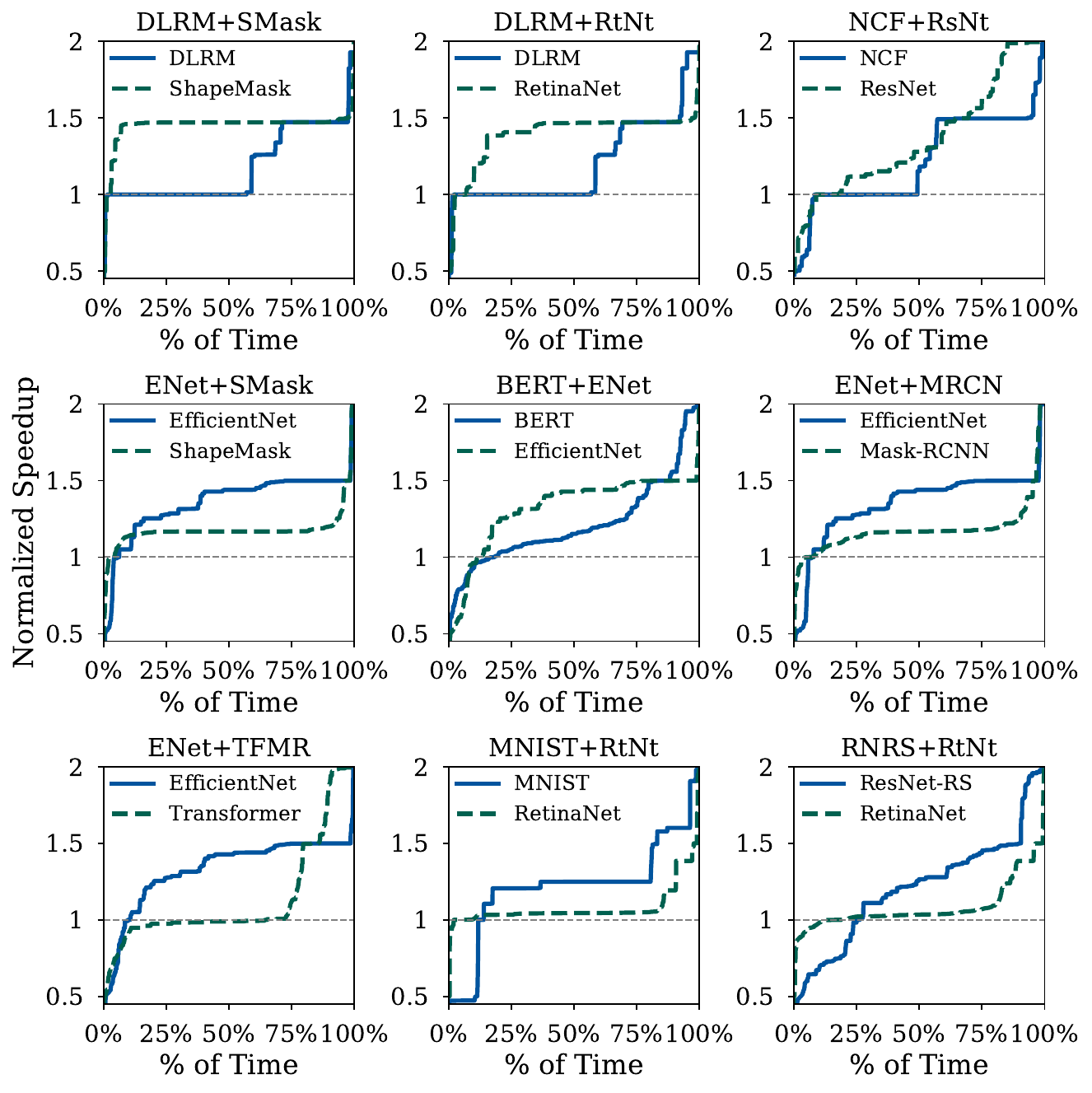}
    \caption{Benefit breakdown of \saname{}/\vuname{} harvesting. Y-axis is the speedup of \pname{} over \nohar{}. X-axis is the percentage of time when executing on \nohar{}. The curves below $Y=1$ indicate the slowdown of operators due to interference. The curves above $Y=1$ indicate the speedup due to harvesting.}
    \label{fig:eval_op_lat_cdf}
\end{figure}

\begin{table}[t]
	\centering
	\caption{The harvesting overhead in each workload, quantified by how much time a workload is blocked due to being harvested over the end-to-end execution time of the workload. ``$<$0.01\%'' means the overhead is smaller than 0.01\%, which rounds to 0 when we only preserve two decimals. For all workloads, the overhead of being harvested is completely outweighed by the benefit of harvesting.}
         \label{tab:eval_harvest_overhead}
	\begin{tabular}{|c|c|c|c|}
		\hline
            \textbf{Collocated Workloads} & \multicolumn{2}{c|}{\textbf{Overhead}} \\\cline{2-3}
		  \textbf{(W1+W2)} & \textbf{W1} & \textbf{W2} \\\hline
		DLRM+SMask & 2.47\% & 0.01\% \\
		DLRM+RtNt & 2.54\% & $<$0.01\% \\
		NCF+RsNt & 6.16\% & $<$0.01\% \\
		ENet+SMask & 5.31\% & 1.12\% \\
		BERT+ENet & $<$0.01\% & 5.54\% \\
		ENet+MRCN & 5.17\% & 1.00\% \\
		ENet+TFMR & 5.61\% & 0.15\% \\
		MNIST+RtNt & 10.63\% & 1.74\% \\
		RNRS+RtNt & 7.33\% & 2.21\% \\
		\hline
	\end{tabular}
\end{table}

\begin{figure*}[t]
    \centering
    \includegraphics[width=\linewidth]{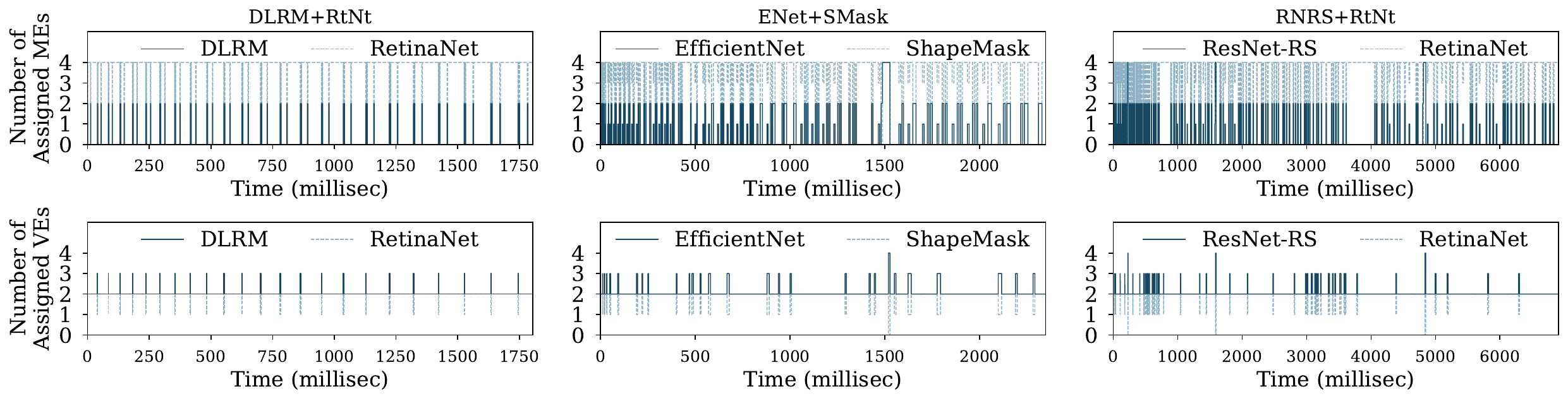}
    \caption{Breakdown of the number of assigned \saname{}s/\vuname{}s over time for different DNN workload combinations.}
    \label{fig:eval_util_vs_time}
\end{figure*}

\vspace{0.2em}
\noindent\textbf{Throughput.}
\Cref{fig:eval_throughput} shows the throughput of the collocated workloads.
When the \saname{}/\vuname{} contention is low, both \ten{} and \pname{} improve the throughput significantly over \pmt{} (by $1.58\times$ and $1.62\times$ on average), 
as the major benefit comes from overlapping the execution of \saname{}-intensive operators and \vuname{}-intensive operators.
When the \saname{}/\vuname{} contention is high, \pname{} improves the throughput of DNN workloads over \ten{} by up to 1.41$\times$, since \pname{} offers more flexibility for dynamic \saname{}/\vuname{} scheduling with \utop{}s, as discussed above.

\subsection{Resource Utilization Improvement}\label{sec:eval_util}

We show the utilization of the \saname{}s and \vuname{}s on the NPU core in \Cref{fig:eval_util}. 
With dynamic operator scheduling, \pname{} improves the \saname{} and \vuname{} utilization by 1.26$\times$ and 1.2$\times$ over \pmt{} on average.
For some workload pairs, \pname{} achieves slightly better utilization than \ten{}, since \pname{} has less preemption 
overhead with \utop{} scheduling. Specifically, \ten{} needs to preempt the entire operator from all \saname{}s, 
while \pname{} only preempts the \utop{}s on the harvested \saname{}s, such that the remaining \utop{}s can continue execution.

\subsection{Benefit Breakdown of \saname{}/\vuname{} Harvesting}\label{sec:eval_breakdown}


To better understand the benefit and overhead of harvesting, we trace the speedup of 
each operator in \pname{} over \nohar{}, and show the impact in~\Cref{fig:eval_op_lat_cdf}.

For workload pairs with low \saname{}/\vuname{} contention (the first row in~\Cref{fig:eval_op_lat_cdf}), most operators achieve at least 1.5$\times$ speedup by harvesting unused compute units from the collocated vNPU with negligible performance interference.
For workload pairs with high \saname{}/\vuname{} contention (the last row in~\Cref{fig:eval_op_lat_cdf}), 
harvesting causes performance degradation for some operators. 
Harvesting may incur extra power and performance overhead (3.12\% on average) when an operator is blocked due to being harvested. We summarize the harvesting overhead in \mbox{\Cref{tab:eval_harvest_overhead}}.
Although harvesting causes slowdowns for some operators,
the overall speedup of the workload still outweighs the slowdowns.



To visualize the behavior of \pname{}'s dynamic \saname{}/\vuname{} scheduling, we trace the number of \saname{}s and \vuname{}s assigned to each collocated workload at runtime in~\Cref{fig:eval_util_vs_time}.
As the \saname{}/\vuname{} demands of the workloads vary across time, the \saname{}-intensive workload (e.g., \texttt{RetinaNet} and \texttt{ShapeMask}) 
attempts to harvest the unused \saname{}s from the collocated workload. 
The \vuname{}s are harvested similarly.


\subsection{Impact of Varying \saname{}s and \vuname{}s}\label{sec:eval_sens_me_ve}

\begin{figure}[t]
    \centering
    \includegraphics[width=\linewidth]{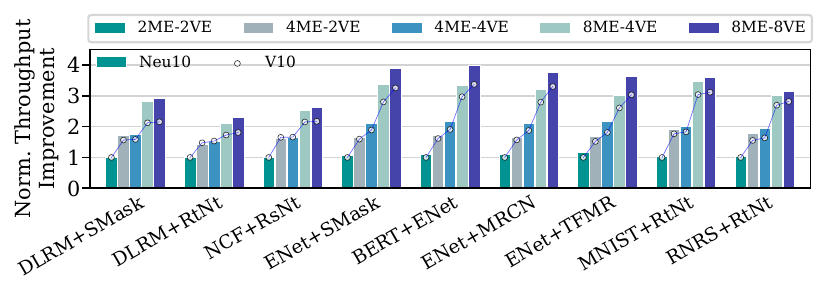}
    \caption{Throughput improvement of \pname{} with varying numbers of \saname{}s and \vuname{}s over \ten{} with 2 \saname{}s and 2 \vuname{}s.}
    \label{fig:eval_vary_me_ve}
\end{figure}

To show \pname{}'s benefits on different hardware configurations, we vary the numbers of \saname{}s and \vuname{}s 
on the physical NPU core and evenly partition the core between the two collocated vNPUs. We compare \pname{} with \ten{}, as \ten{} has 
fine-grained preemption, which serves as the most competitive baseline. 
We show the throughput in~\Cref{fig:eval_vary_me_ve}.
With more \saname{}s/\vuname{}s, \pname{} brings more benefits, 
since there is more flexibility for dynamic \saname{}/\vuname{} scheduling.
With more \saname{}s/\vuname{}s, it is more likely that an operator cannot fully exploit all \saname{}/\vuname{}s. Therefore, the benefit of \utop{}-level scheduling and harvesting becomes more obvious.




\subsection{Impact of Varying Memory Bandwidth}\label{sec:eval_sens_hbm_bw}

\begin{figure}[t]
    \centering
    \includegraphics[width=\linewidth]{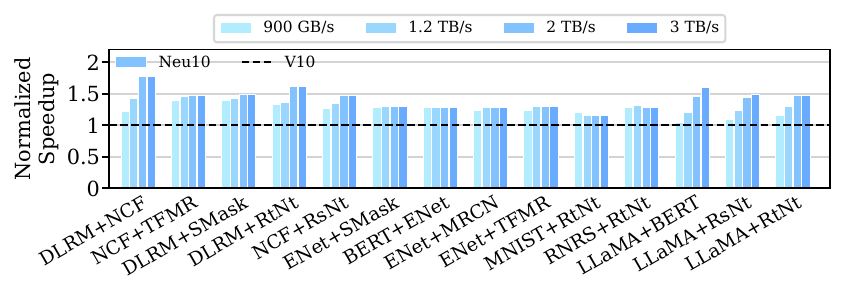}
    \caption{Throughput improvement of {\pname{}} with varying HBM bandwidth (normalized to \ten{}).}
    \label{fig:eval_vary_hbm_bw}
\end{figure}

We show \pname{}'s performance under different HBM bandwidth configurations in \Cref{fig:eval_vary_hbm_bw}.
In most cases, \pname{} achieves similar throughput benefits. 
This is because many ML inference workloads suffer from the \saname{}/\vuname{} contention rather than HBM bandwidth contention (see~\Cref{fig:hbm_bw_util_ops}).
To understand the impact of memory bandwidth contention, we collocate two memory-intensive workloads (i.e., \texttt{DLRM+NCF} and \texttt{NCF+TFMR}).
Even with low available memory bandwidth (e.g., 900 GB/s), \pname{} still outperforms the time sharing-based scheme \ten{}.
With more available bandwidth, \pname{} brings more benefits for memory-intensive workloads, 
since higher bandwidth helps alleviate memory contention. 

\begin{figure}[t]
    \centering
    \includegraphics[width=\linewidth]{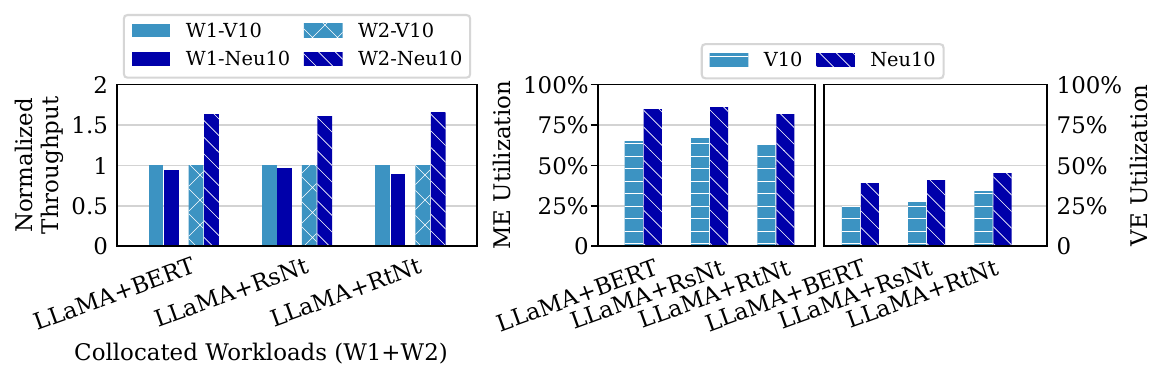}
    \caption{Performance of collocating LLM (LLaMA-13B with batch size 8 and input sequence length 512) and other models in \pname{}.}
    \label{fig:eval_llm}
\end{figure}

For memory-intensive workloads, \pname{} enables them to be collocated with compute-intensive workloads following existing workload collocation approaches~\cite{v10:isca23,prophet:asplos2017,heracles:isca2015}, which helps cloud platforms better utilize both compute and memory resources.
As a case study, we collocate a memory bandwidth-intensive LLM inference workload, LLaMA2-13B~\cite{llama2} (\texttt{LLaMA}), with compute-intensive workloads (i.e., \texttt{BERT}, \texttt{RsNt}, and \texttt{RtNt}).
As shown in \Cref{fig:eval_llm}, 
with \ten{}, when LLaMA temporarily occupies all \saname{}s/\vuname{}s, it underutilizes the \saname{}s/\vuname{}s since the execution is bounded by memory bandwidth. However, the underutilized \saname{}s cannot be harvested by the collocated workload due to the temporal sharing mechanism.
{\pname{}} enables the spatial sharing of the \saname{}s/\vuname{}s, so the collocated workload can harvest the spare \saname{}s/\vuname{}s for throughput improvement (by up to 1.6$\times$).
Meanwhile, LLaMA suffers from negligible overhead while using fewer \saname{}s/\vuname{}s.
As hardware vendors continue to scale the HBM bandwidth, \pname{} will bring more benefits because of the alleviation of memory contention. Note that the owners of LLM inference service can follow the pay-as-you-go model to allocate multiple vNPUs with large memory. The vNPU abstraction of \pname{} offers the flexibility for resource allocation while enabling dynamic scheduling. 

\section{Related Work}\label{sec:related_work}

%

\noindent
\textbf{System virtualization for accelerators.}
As we employ hardware accelerators for ML services, cloud platforms prefer to virtualize them for improved 
resource utilization~\cite{fpgacloud,gpucloud,micropreempt:osdi2022,vital:asplos20,synergy:asplos21,mlvital:asplos21,
amorphos:osdi18}. Prior studies have investigated virtualization techniques 
for GPUs~\cite{vgpu:nvidia, mig:nvidia, preemptgpu:isca14, preemptsimt:sc16} and FPGAs~\cite{amorphos:osdi18, 
optimus:asplos20, vital:asplos20, synergy:asplos21, mlvital:asplos21}. Unfortunately, these techniques cannot be 
directly applied to NPUs, as they target different architectures. 
AvA~\cite{ava:asplos20} investigates hypervisor interposition techniques for virtualizing accelerators. However, they do not focus on improving the resource utilization.
To the best of our knowledge, \pname{} 
is the first to investigate the system and architectural techniques for NPU virtualization. 

While different tensor processors have been developed recently~\cite{ascend910:huawei:hotchips31,hanguang:ali:hotchips32,aws_inferentia,tpucloud}, most of them have specialized matrix engines and generic vector engines,  given the continuing trend that DNN computations are dominated by these operations.
As the imbalanced {\saname{}}/{\vuname{}} demands are intrinsic to DNN workloads ($\S$\ref{sec:npu_util_study}), these processors also suffer from resource underutilization.
The design of \pname{} can be adapted to virtualize these accelerators for utilization improvement.

\vspace{0.2em}
\noindent
\textbf{Accelerator resource sharing and scheduling.}
There have been various techniques~\cite{micropreempt:osdi2022,multiteantgpu:osdi2022,baymax:asplos2016, prophet:asplos2017,amorphos:osdi18, optimus:asplos20, vital:asplos20, synergy:asplos21,
mlvital:asplos21,neucloud:hotos23,vnpu:hotinfra23} for supporting multi-tenant workloads on accelerators. PREMA~\cite{prema:hpca20} proposed a preemptive scheduling mechanism, but it causes high context-switch overhead.
Prior studies~\cite{aimt:isca20,layerweaver:hpca21,v10:isca23} investigated the imbalance of compute units and memory.
Planaria~\cite{planaria:micro20} studied the spatial underutilization of systolic arrays. 
V10~\cite{v10:isca23} enabled fine-grained preemption. 
There are also software techniques to fuse DNN workloads at graph level~\cite{modelfusion:iccad2021,gmorph:eurosys24}. However, they force two DNN inference tasks to launch together, which cannot work for the unpredictable incoming requests in the cloud.
None of them systematically enables NPU virtualization.
\pname{} addressed the systems and architectural challenges of NPU virtualization. 

As the resource demand of a DNN workload changes drastically over time (see \S\ref{sec:npu_util_study}), a static resource allocation is insufficient.
\pname{} proposes NeuISA and enables dynamic scheduling to mitigate the underutilization, 
it is orthogonal to the higher-level workload collocation techniques.

\vspace{0.2em}
\noindent
\textbf{Architectural support for virtualization.}
Prior studies have proven that architectural techniques are effective for facilitating system virtualization~\cite{xen:sosp2003}, including Intel VT-x and AMD SVM for CPU virtualization~\cite{adams:asplos2006}, Intel EPT~\cite{intel-ept} and AMD NPT for memory 
address translation~\cite{ravi:asplos2008}, and SR-IOV for I/O virtualization~\cite{ruivo:ccgrid2014, sr-iov}. 
Similarly, NPU virtualization also needs architectural support.
In this work, we identify the unique architectural challenges (see $\S$\ref{sec:intro}) with NPU virtualization, and present the corresponding ISA extension and architectural supports for enabling fine-grained NPU virtualization. 

\section{Conclusion}
\label{sec:conclusion}
We identify the key challenges of virtualizing NPUs for cloud platforms, 
including the need for fine-grained system abstraction and resource scheduling
and the necessity of architectural support. We present a holistic solution \pname{} for enabling 
NPU virtualization.
It improves both NPU utilization and performance isolation for multi-tenant ML services.  

\section*{Acknowledgements}

We thank the anonymous reviewers for their helpful comments and feedback.
We thank Haoyang Zhang for his insightful discussion on the \isa design.
This work was partially supported by NSF grant CCF-1919044, NSF CAREER Award CNS-2144796, and the Hybrid Cloud and AI program at the IBM-Illinois Discovery Accelerator Institute (IIDAI). 

\bibliographystyle{IEEEtranS}
\bibliography{IEEEfull,references}

\begin{thebibliography}{10}
\providecommand{\url}[1]{#1}
\csname url@samestyle\endcsname
\providecommand{\newblock}{\relax}
\providecommand{\bibinfo}[2]{#2}
\providecommand{\BIBentrySTDinterwordspacing}{\spaceskip=0pt\relax}
\providecommand{\BIBentryALTinterwordstretchfactor}{4}
\providecommand{\BIBentryALTinterwordspacing}{\spaceskip=\fontdimen2\font plus
\BIBentryALTinterwordstretchfactor\fontdimen3\font minus
  \fontdimen4\font\relax}
\providecommand{\BIBforeignlanguage}[2]{{%
\expandafter\ifx\csname l@#1\endcsname\relax
\typeout{** WARNING: IEEEtranS.bst: No hyphenation pattern has been}%
\typeout{** loaded for the language `#1'. Using the pattern for}%
\typeout{** the default language instead.}%
\else
\language=\csname l@#1\endcsname
\fi
#2}}
\providecommand{\BIBdecl}{\relax}
\BIBdecl

\bibitem{freepdk15}
\BIBentryALTinterwordspacing
``{FreePDK15}.'' [Online]. Available: \url{https://eda.ncsu.edu/freepdk15/}
\BIBentrySTDinterwordspacing

\bibitem{memory_segmentation:wikipedia}
\BIBentryALTinterwordspacing
``Memory segmentation.'' [Online]. Available:
  \url{https://en.wikipedia.org/wiki/Memory_segmentation}
\BIBentrySTDinterwordspacing

\bibitem{adams:asplos2006}
K.~Adams and O.~Agesen, ``A comparison of software and hardware techniques for
  x86 virtualization,'' in \emph{Proceedings of the 12th International
  Conference on Architectural Support for Programming Languages and Operating
  Systems (ASPLOS'06)}, San Jose, CA, USA, 2006.

\bibitem{mlaas:industry}
\BIBentryALTinterwordspacing
Altexsoft, ``{Comparing Machine Learning as a Service: Amazon, Microsoft Azure,
  Google Cloud AI, IBM Watson},'' 2021. [Online]. Available:
  \url{https://www.altexsoft.com/blog/datascience/comparing-machine-learning-as-a-service-amazon-microsoft-azure-google-cloud-ai-ibm-watson/}
\BIBentrySTDinterwordspacing

\bibitem{amd:aiengine}
\BIBentryALTinterwordspacing
AMD, ``{AI Engine: Meeting the Compute Demands of Next-Generation
  Applications},'' 2023. [Online]. Available:
  \url{https://www.xilinx.com/products/technology/ai-engine.html}
\BIBentrySTDinterwordspacing

\bibitem{fpgacloud}
\BIBentryALTinterwordspacing
AWS, ``{Amazon EC2 F1 Instances},'' 2022. [Online]. Available:
  \url{https://aws.amazon.com/ec2/instance-types/f1/}
\BIBentrySTDinterwordspacing

\bibitem{mlaas:aws}
\BIBentryALTinterwordspacing
A.~AWS, ``{Machine Learning on AWS Innovate faster with the most comprehensive
  set of AI and ML services},'' 2022. [Online]. Available:
  \url{https://aws.amazon.com/machine-learning/}
\BIBentrySTDinterwordspacing

\bibitem{aws_inferentia}
\BIBentryALTinterwordspacing
A.~AWS, ``Aws inferentia,'' 2023. [Online]. Available:
  \url{https://aws.amazon.com/machine-learning/inferentia/}
\BIBentrySTDinterwordspacing

\bibitem{aimt:isca20}
E.~Baek, D.~Kwon, and J.~Kim, ``A multi-neural network acceleration
  architecture,'' in \emph{Proceedings of the ACM/IEEE 47th Annual
  International Symposium on Computer Architecture (ISCA'20)}, Virtual Event,
  2020.

\bibitem{xen:sosp2003}
P.~Barham, B.~Dragovic, K.~Fraser, S.~Hand, T.~Harris, A.~Ho, R.~Neugebauer,
  I.~Pratt, and A.~Warfield, ``Xen and the art of virtualization,'' in
  \emph{Proceedings of the Nineteenth ACM Symposium on Operating Systems
  Principles (SOSP'03)}, Bolton Landing, NY, USA, 2003.

\bibitem{ravi:asplos2008}
R.~Bhargava, B.~Serebrin, F.~Spadini, and S.~Manne, ``Accelerating
  two-dimensional page walks for virtualized systems,'' in \emph{Proceedings of
  the 13th International Conference on Architectural Support for Programming
  Languages and Operating Systems (ASPLOS'08)}, Seattle, WA, USA, 2008.

\bibitem{prophet:asplos2017}
Q.~Chen, H.~Yang, M.~Guo, R.~S. Kannan, J.~Mars, and L.~Tang, ``Prophet:
  Precise qos prediction on non-preemptive accelerators to improve utilization
  in warehouse-scale computers,'' in \emph{Proceedings of the Twenty-Second
  International Conference on Architectural Support for Programming Languages
  and Operating Systems (ASPLOS'17)}, Xi'an, China, 2017.

\bibitem{baymax:asplos2016}
Q.~Chen, H.~Yang, J.~Mars, and L.~Tang, ``Baymax: Qos awareness and increased
  utilization for non-preemptive accelerators in warehouse scale computers,''
  in \emph{Proceedings of the Twenty-First International Conference on
  Architectural Support for Programming Languages and Operating Systems
  (ASPLOS'16)}, Atlanta, GA, 2016.

\bibitem{tvm}
T.~Chen, T.~Moreau, Z.~Jiang, L.~Zheng, E.~Yan, H.~Shen, M.~Cowan, L.~Wang,
  Y.~Hu, L.~Ceze, C.~Guestrin, and A.~Krishnamurthy, ``{{TVM}: An Automated
  {End-to-End} Optimizing Compiler for Deep Learning},'' in \emph{Proceedings
  of the 13th USENIX Symposium on Operating Systems Design and Implementation
  (OSDI'18)}, Carlsbad, CA, 2018.

\bibitem{diannao}
T.~Chen, Z.~Du, N.~Sun, J.~Wang, C.~Wu, Y.~Chen, and O.~Temam, ``{DianNao: A
  Small-Footprint High-Throughput Accelerator for Ubiquitous
  Machine-Learning},'' in \emph{Proceedings of the 20th International
  Conference on Architectural Support for Programming Languages and Operating
  Systems (ASPLOS'14)}, Salt Lake City, UT, 2014.

\bibitem{prema:hpca20}
Y.~Choi and M.~Rhu, ``{PREMA}: A predictive multi-task scheduling algorithm for
  preemptible neural processing units,'' in \emph{Proceedings of the 2020 IEEE
  International Symposium on High Performance Computer Architecture (HPCA'20)},
  San Diego, CA, USA, 2020.

\bibitem{brainwave}
E.~Chung, J.~Fowers, K.~Ovtcharov, M.~Papamichael, A.~Caulfield, T.~Massengil,
  M.~Liu, D.~Lo, S.~Alkalay, M.~Haselman, C.~Boehn, O.~Firestein, A.~Forin,
  K.~S. Gatlin, M.~Ghandi, S.~Heil, K.~Holohan, T.~Juhasz, R.~K. Kovvuri,
  S.~Lanka, F.~van Megen, D.~Mukhortov, P.~Patel, S.~Reinhardt, A.~Sapek,
  R.~Seera, B.~Sridharan, L.~Woods, P.~Yi-Xiao, R.~Zhao, and D.~Burger,
  ``{Accelerating Persistent Neural Networks at Datacenter Scale},'' in
  \emph{Proceedings of HotChips'17}, Cupertino, CA, USA, 2017.

\bibitem{ruivo:ccgrid2014}
T.~P.~P. de~Lacerda~Ruivo, G.~B. Altayo, G.~Garzoglio, S.~Timm, H.~W. Kim,
  S.-Y. Noh, and I.~Raicu, ``Exploring infiniband hardware virtualization in
  opennebula towards efficient high-performance computing,'' in
  \emph{Proceedings of the 14th IEEE/ACM International Symposium on Cluster,
  Cloud and Grid Computing (CCGrid'14)}, Chicago, IL, USA, 2014.

\bibitem{planaria:micro20}
S.~Ghodrati, B.~H. Ahn, J.~Kyung~Kim, S.~Kinzer, B.~R. Yatham, N.~Alla,
  H.~Sharma, M.~Alian, E.~Ebrahimi, N.~S. Kim, C.~Young, and H.~Esmaeilzadeh,
  ``Planaria: Dynamic architecture fission for spatial multi-tenant
  acceleration of deep neural networks,'' in \emph{Proceedings of the 53rd
  Annual IEEE/ACM International Symposium on Microarchitecture (MICRO'20)},
  Virtual Event, 2020.

\bibitem{cloudtpu:google}
\BIBentryALTinterwordspacing
Google, ``System architecture - cloud {TPU},'' 2022. [Online]. Available:
  \url{https://cloud.google.com/tpu/docs/system-architecture-tpu-vm}
\BIBentrySTDinterwordspacing

\bibitem{tensorflow}
\BIBentryALTinterwordspacing
Google, ``{Create production-grade machine learning models with TensorFlow},''
  2023. [Online]. Available: \url{https://www.tensorflow.org/}
\BIBentrySTDinterwordspacing

\bibitem{tpu_supported_models}
\BIBentryALTinterwordspacing
Google, ``Supported reference models,'' 2023. [Online]. Available:
  \url{https://cloud.google.com/tpu/docs/tutorials/supported-models}
\BIBentrySTDinterwordspacing

\bibitem{xla}
\BIBentryALTinterwordspacing
Google, ``{XLA}: {O}ptimizing {C}ompiler for {M}achine {L}earning,'' 2023.
  [Online]. Available: \url{https://www.tensorflow.org/xla}
\BIBentrySTDinterwordspacing

\bibitem{ipu:graphcore}
\BIBentryALTinterwordspacing
Graphcore, ``Graphcore {IPU} overview,'' 2022. [Online]. Available:
  \url{https://www.graphcore.ai/products/ipu}
\BIBentrySTDinterwordspacing

\bibitem{tenstorrent}
\BIBentryALTinterwordspacing
L.~Gwennap, ``Tenstorrent scales ai performance: New multicore architecture
  leads in data-center power efficiency,'' 2020. [Online]. Available:
  \url{https://www.linleygroup.com/mpr/article.php?id=12287}
\BIBentrySTDinterwordspacing

\bibitem{micropreempt:osdi2022}
M.~Han, H.~Zhang, R.~Chen, and H.~Chen, ``Microsecond-scale preemption for
  concurrent {GPU-accelerated} {DNN} inferences,'' in \emph{Proccedings of the
  16th USENIX Symposium on Operating Systems Design and Implementation
  (OSDI'22)}, Carlsbad, CA, USA, 2022.

\bibitem{swapadvisor:asplos20}
C.-C. Huang, G.~Jin, and J.~Li, ``Swapadvisor: Pushing deep learning beyond the
  gpu memory limit via smart swapping,'' in \emph{Proceedings of the 25th
  International Conference on Architectural Support for Programming Languages
  and Operating Systems (ASPLOS'20)}, Lausanne, Switzerland, 2020.

\bibitem{aichips:industry}
\BIBentryALTinterwordspacing
J.~Hui, ``{AI Chips Technology Trends and Landscapes (Mobile SoC, Intel, Asian
  AI Chips, Low-Power Inference Chips)},'' 2020. [Online]. Available:
  \url{https://jonathan-hui.medium.com/ai-chips-technology-trends-landscape-mobile-soc-intel-asian-ai-chips-low-power-inference-4db701dbe85d}
\BIBentrySTDinterwordspacing

\bibitem{vfio-mdev}
\BIBentryALTinterwordspacing
N.~Jia and K.~Wankhede, ``Vfio mediated devices,'' 2023. [Online]. Available:
  \url{https://docs.kernel.org/driver-api/vfio-mediated-device.html}
\BIBentrySTDinterwordspacing

\bibitem{hanguang:ali:hotchips32}
Y.~Jiao, L.~Han, and X.~Long, ``Hanguang 800 npu – the ultimate ai inference
  solution for data centers,'' in \emph{2020 IEEE Hot Chips 32 Symposium
  (HCS)}, Palo Alto, CA, USA, 2020.

\bibitem{tpuarch:google:commACM20}
N.~P. Jouppi, D.~H. Yoon, G.~Kurian, S.~Li, N.~Patil, J.~Laudon, C.~Young, and
  D.~Patterson, ``A domain-specific supercomputer for training deep neural
  networks,'' \emph{Commun. ACM}, vol.~63, no.~7, June 2020.

\bibitem{tpu}
N.~P. Jouppi, C.~Young, N.~Patil, D.~Patterson, G.~Agrawal, R.~Bajwa, S.~Bates,
  S.~Bhatia, N.~Boden, A.~Borchers, R.~Boyle, P.-l. Cantin, C.~Chao, C.~Clark,
  J.~Coriell, M.~Daley, M.~Dau, J.~Dean, B.~Gelb, T.~V. Ghaemmaghami,
  R.~Gottipati, W.~Gulland, R.~Hagmann, C.~R. Ho, D.~Hogberg, J.~Hu, R.~Hundt,
  D.~Hurt, J.~Ibarz, A.~Jaffey, A.~Jaworski, A.~Kaplan, H.~Khaitan,
  D.~Killebrew, A.~Koch, N.~Kumar, S.~Lacy, J.~Laudon, J.~Law, D.~Le, C.~Leary,
  Z.~Liu, K.~Lucke, A.~Lundin, G.~MacKean, A.~Maggiore, M.~Mahony, K.~Miller,
  R.~Nagarajan, R.~Narayanaswami, R.~Ni, K.~Nix, T.~Norrie, M.~Omernick,
  N.~Penukonda, A.~Phelps, J.~Ross, M.~Ross, A.~Salek, E.~Samadiani, C.~Severn,
  G.~Sizikov, M.~Snelham, J.~Souter, D.~Steinberg, A.~Swing, M.~Tan,
  G.~Thorson, B.~Tian, H.~Toma, E.~Tuttle, V.~Vasudevan, R.~Walter, W.~Wang,
  E.~Wilcox, and D.~H. Yoon, ``{In-Datacenter Performance Analysis of a Tensor
  Processing Unit},'' in \emph{Proceedings of the 44th International Symposium
  on Computer Architecture (ISCA'17)}, Toronto, Canada, 2017.

\bibitem{amorphos:osdi18}
A.~Khawaja, J.~Landgraf, R.~Prakash, M.~Wei, E.~Schkufza, and C.~J. Rossbach,
  ``Sharing, protection, and compatibility for reconfigurable fabric with
  {AmorphOS},'' in \emph{13th USENIX Symposium on Operating Systems Design and
  Implementation (OSDI'18)}, Carlsbad, CA, USA, 2018.

\bibitem{synergy:asplos21}
J.~Landgraf, T.~Yang, W.~Lin, C.~J. Rossbach, and E.~Schkufza,
  ``Compiler-driven fpga virtualization with synergy,'' in \emph{Proceedings of
  the 26th ACM International Conference on Architectural Support for
  Programming Languages and Operating Systems (ASPLOS'21)}, Virtual Event,
  2021.

\bibitem{ascend910:huawei:hotchips31}
H.~Liao, J.~Tu, J.~Xia, and X.~Zhou, ``Davinci: A scalable architecture for
  neural network computing,'' in \emph{2019 IEEE Hot Chips 31 Symposium (HCS)},
  Los Alamitos, CA, USA, 2019.

\bibitem{preemptsimt:sc16}
Z.~Lin, L.~Nyland, and H.~Zhou, ``Enabling efficient preemption for simt
  architectures with lightweight context switching,'' in \emph{Proceedings of
  the International Conference for High Performance Computing, Networking,
  Storage and Analysis (SC'16)}, Salt Lake City, UT, USA, 2016.

\bibitem{heracles:isca2015}
D.~Lo, L.~Cheng, R.~Govindaraju, P.~Ranganathan, and C.~Kozyrakis, ``Heracles:
  Improving resource efficiency at scale,'' in \emph{Proceedings of the 42nd
  Annual International Symposium on Computer Architecture (ISCA'15)}, Portland,
  OR, USA, 2015.

\bibitem{optimus:asplos20}
J.~Ma, G.~Zuo, K.~Loughlin, X.~Cheng, Y.~Liu, A.~M. Eneyew, Z.~Qi, and
  B.~Kasikci, ``A hypervisor for shared-memory fpga platforms,'' in
  \emph{Proceedings of the 25th International Conference on Architectural
  Support for Programming Languages and Operating Systems (ASPLOS'20)},
  Lausanne, Switzerland, 2020.

\bibitem{bubbleup:micro11}
J.~Mars, L.~Tang, R.~Hundt, K.~Skadron, and M.~L. Soffa, ``Bubble-up:
  Increasing utilization in modern warehouse scale computers via sensible
  co-locations,'' in \emph{Proceedings of the 44th Annual IEEE/ACM
  International Symposium on Microarchitecture (MICRO'11)}, Porto Alegre,
  Brazil, 2011.

\bibitem{multiteantgpu:osdi2022}
J.~Mohan, A.~Phanishayee, J.~Kulkarni, and V.~Chidambaram, ``Looking beyond
  {GPUs} for {DNN} scheduling on {Multi-Tenant} clusters,'' in
  \emph{Proceedings of the 16th USENIX Symposium on Operating Systems Design
  and Implementation (OSDI'22)}, Carlsbad, CA, USA, 2022.

\bibitem{mig:nvidia}
\BIBentryALTinterwordspacing
Nvidia, ``{M}ulti-{I}nstance {GPU} user guide,'' 2022. [Online]. Available:
  \url{https://docs.nvidia.com/datacenter/tesla/mig-user-guide/}
\BIBentrySTDinterwordspacing

\bibitem{vgpu:nvidia}
\BIBentryALTinterwordspacing
Nvidia, ``Virtual {GPU} software user guide,'' 2022. [Online]. Available:
  \url{https://docs.nvidia.com/grid/latest/grid-vgpu-user-guide/}
\BIBentrySTDinterwordspacing

\bibitem{layerweaver:hpca21}
Y.~H. Oh, S.~Kim, Y.~Jin, S.~Son, J.~Bae, J.~Lee, Y.~Park, D.~U. Kim, T.~J.
  Ham, and J.~W. Lee, ``Layerweaver: Maximizing resource utilization of neural
  processing units via layer-wise scheduling,'' in \emph{2021 IEEE
  International Symposium on High-Performance Computer Architecture (HPCA'21)},
  Seoul, Korea, 2021.

\bibitem{mlaas:intro}
\BIBentryALTinterwordspacing
E.~Onose, ``Machine learning as a service: What it is, when to use it and what
  are the best tools out there,'' 2022. [Online]. Available:
  \url{https://neptune.ai/blog/machine-learning-as-a-service-what-it-is-when-to-use-it-and-what-are-the-best-tools-out-there}
\BIBentrySTDinterwordspacing

\bibitem{PyTorch}
A.~Paszke, S.~Gross, S.~Chintala, G.~Chanan, E.~Yang, Z.~DeVito, Z.~Lin,
  A.~Desmaison, L.~Antiga, and A.~Lerer, ``{Automatic Differentiation in
  PyTorch},'' in \emph{Proceedings of the 30th International Conference on
  Neural Information Processing Systems (NIPS'17)}, Long Beach, CA, USA, 2017.

\bibitem{reddi2019mlperf}
V.~J. Reddi, C.~Cheng, D.~Kanter, P.~Mattson, G.~Schmuelling, C.-J. Wu,
  B.~Anderson, M.~Breughe, M.~Charlebois, W.~Chou, R.~Chukka, C.~Coleman,
  S.~Davis, P.~Deng, G.~Diamos, J.~Duke, D.~Fick, J.~S. Gardner, I.~Hubara,
  S.~Idgunji, T.~B. Jablin, J.~Jiao, T.~S. John, P.~Kanwar, D.~Lee, J.~Liao,
  A.~Lokhmotov, F.~Massa, P.~Meng, P.~Micikevicius, C.~Osborne, G.~Pekhimenko,
  A.~T.~R. Rajan, D.~Sequeira, A.~Sirasao, F.~Sun, H.~Tang, M.~Thomson, F.~Wei,
  E.~Wu, L.~Xu, K.~Yamada, B.~Yu, G.~Yuan, A.~Zhong, P.~Zhang, and Y.~Zhou,
  ``Mlperf inference benchmark,'' 2020.

\bibitem{tpucloud}
\BIBentryALTinterwordspacing
RUN:AI, ``{Google TPU Architecture and Performance Best Practices},'' 2022.
  [Online]. Available:
  \url{https://www.run.ai/guides/cloud-deep-learning/google-tpu}
\BIBentrySTDinterwordspacing

\bibitem{pay-as-you-go}
\BIBentryALTinterwordspacing
{Stephen J. Bigelow}, ``pay-as-you-go cloud computing ({PAYG} cloud
  computing),'' 2022. [Online]. Available:
  \url{https://www.techtarget.com/searchstorage/definition/pay-as-you-go-cloud-computing-PAYG-cloud-computing}
\BIBentrySTDinterwordspacing

\bibitem{preemptgpu:isca14}
I.~Tanasic, I.~Gelado, J.~Cabezas, A.~Ramirez, N.~Navarro, and M.~Valero,
  ``Enabling preemptive multiprogramming on gpus,'' in \emph{2014 ACM/IEEE 41st
  International Symposium on Computer Architecture (ISCA)}, 2014.

\bibitem{kubevirt}
\BIBentryALTinterwordspacing
{The KubeVirt Contributors}, ``Kubevirt.io,'' 2023. [Online]. Available:
  \url{https://kubevirt.io/}
\BIBentrySTDinterwordspacing

\bibitem{llama2}
H.~Touvron, L.~Martin, K.~Stone, P.~Albert, A.~Almahairi, Y.~Babaei,
  N.~Bashlykov, S.~Batra, P.~Bhargava, S.~Bhosale, D.~Bikel, L.~Blecher, C.~C.
  Ferrer, M.~Chen, G.~Cucurull, D.~Esiobu, J.~Fernandes, J.~Fu, W.~Fu,
  B.~Fuller, C.~Gao, V.~Goswami, N.~Goyal, A.~Hartshorn, S.~Hosseini, R.~Hou,
  H.~Inan, M.~Kardas, V.~Kerkez, M.~Khabsa, I.~Kloumann, A.~Korenev, P.~S.
  Koura, M.-A. Lachaux, T.~Lavril, J.~Lee, D.~Liskovich, Y.~Lu, Y.~Mao,
  X.~Martinet, T.~Mihaylov, P.~Mishra, I.~Molybog, Y.~Nie, A.~Poulton,
  J.~Reizenstein, R.~Rungta, K.~Saladi, A.~Schelten, R.~Silva, E.~M. Smith,
  R.~Subramanian, X.~E. Tan, B.~Tang, R.~Taylor, A.~Williams, J.~X. Kuan,
  P.~Xu, Z.~Yan, I.~Zarov, Y.~Zhang, A.~Fan, M.~Kambadur, S.~Narang,
  A.~Rodriguez, R.~Stojnic, S.~Edunov, and T.~Scialom, ``Llama 2: Open
  foundation and fine-tuned chat models,'' 2023.

\bibitem{tpuv5p}
\BIBentryALTinterwordspacing
A.~Vahdat and M.~Lohmeyer, ``Enabling next-generation ai workloads: Announcing
  tpu v5p and ai hypercomputer,'' 2023. [Online]. Available:
  \url{https://cloud.google.com/blog/products/ai-machine-learning/introducing-cloud-tpu-v5p-and-ai-hypercomputer}
\BIBentrySTDinterwordspacing

\bibitem{intel-ept}
\BIBentryALTinterwordspacing
VMWare, ``{Performance Evaluation of Intel EPT Hardware Assist},'' 2009.
  [Online]. Available:
  \url{https://www.vmware.com/pdf/Perf_ESX_Intel-EPT-eval.pdf}
\BIBentrySTDinterwordspacing

\bibitem{sr-iov}
\BIBentryALTinterwordspacing
VMWare, ``{vSphere Networking},'' 2009. [Online]. Available:
  \url{https://docs.vmware.com/en/VMware-vSphere/8.0/vsphere-esxi-vcenter-802-networking-guide.pdf}
\BIBentrySTDinterwordspacing

\bibitem{gpucloud}
\BIBentryALTinterwordspacing
K.~Wiggers, ``Microsoft and nvidia team up to build new azure-hosted ai
  supercomputer,'' 2022. [Online]. Available:
  \url{https://techcrunch.com/2022/11/16/microsoft-and-nvidia-team-up-to-build-new-azure-hosted-ai-supercomputer/}
\BIBentrySTDinterwordspacing

\bibitem{wolfram_alpha}
\BIBentryALTinterwordspacing
{Wolfram Alpha LLC}, ``Wolframalpha: Computational intelligence,'' 2023.
  [Online]. Available: \url{https://www.wolframalpha.com/}
\BIBentrySTDinterwordspacing

\bibitem{neucloud:hotos23}
Y.~Xue, Y.~Liu, and J.~Huang, ``System virtualization for neural processing
  units,'' in \emph{Proceedings of the 19th Workshop on Hot Topics in Operating
  Systems (HotOS'23)}, Providence, RI, USA, 2023.

\bibitem{vnpu:hotinfra23}
Y.~Xue, Y.~Liu, L.~Nai, and J.~Huang, ``Hardware-assisted virtualization for
  neural processing units,'' in \emph{The 1st Workshop on Hot Topics in System
  Infrastructure (HotInfra'23)}, Orlando, FL, USA, 2023.

\bibitem{v10:isca23}
Y.~Xue, Y.~Liu, L.~Nai, and J.~Huang, ``V10: Hardware-assisted npu
  multi-tenancy for improved resource utilization and fairness,'' in
  \emph{Proceedings of the 50th Annual International Symposium on Computer
  Architecture (ISCA'23)}, Orlando, FL, USA, 2023.

\bibitem{gmorph:eurosys24}
Q.~Yang, T.~Yang, M.~Xiang, L.~Zhang, H.~Wang, M.~Serafini, and H.~Guan,
  ``{GMorph}: Accelerating multi-dnn inference via model fusion,'' in
  \emph{Proceedings of the 19th European Conference on Computer Systems
  (EuroSys'24)}, Athens, Greece.

\bibitem{modelfusion:iccad2021}
F.~Yu, S.~Bray, D.~Wang, L.~Shangguan, X.~Tang, C.~Liu, and X.~Chen,
  ``Automated runtime-aware scheduling for multi-tenant dnn inference on gpu,''
  in \emph{2021 IEEE/ACM International Conference On Computer Aided Design
  (ICCAD)}, 2021.

\bibitem{ava:asplos20}
H.~Yu, A.~M. Peters, A.~Akshintala, and C.~J. Rossbach, ``{AvA}: Accelerated
  virtualization of accelerators,'' in \emph{Proceedings of the 25th
  International Conference on Architectural Support for Programming Languages
  and Operating Systems (ASPLOS'20)}, Lausanne, Switzerland, 2020.

\bibitem{vital:asplos20}
Y.~Zha and J.~Li, ``Virtualizing fpgas in the cloud,'' in \emph{Proceedings of
  the 25th International Conference on Architectural Support for Programming
  Languages and Operating Systems (ASPLOS'20)}, Lausanne, Switzerland, 2020.

\bibitem{mlvital:asplos21}
Y.~Zha and J.~Li, ``When application-specific isa meets fpgas: A multi-layer
  virtualization framework for heterogeneous cloud fpgas,'' in
  \emph{Proceedings of the 26th ACM International Conference on Architectural
  Support for Programming Languages and Operating Systems (ASPLOS'21)}, Virtual
  Event, 2021.

\bibitem{g10:micro23}
H.~Zhang, Y.~Zhou, Y.~Xue, Y.~Liu, and J.~Huang, ``G10: Enabling an efficient
  unified gpu memory and storage architecture with smart tensor migrations,''
  in \emph{Proceedings of the 56th Annual IEEE/ACM International Symposium on
  Microarchitecture (MICRO'23)}, Toronto, ON, Canada, 2023.

\bibitem{roller:osdi2022}
H.~Zhu, R.~Wu, Y.~Diao, S.~Ke, H.~Li, C.~Zhang, J.~Xue, L.~Ma, Y.~Xia, W.~Cui,
  F.~Yang, M.~Yang, L.~Zhou, A.~Cidon, and G.~Pekhimenko, ``{ROLLER}: Fast and
  efficient tensor compilation for deep learning,'' in \emph{Proceedings of
  16th USENIX Symposium on Operating Systems Design and Implementation
  (OSDI'22)}, Carlsbad, CA, USA, 2022.

\end{thebibliography}

\end{document}